# Improved Gauss law model and in-medium heavy quarkonium at finite density and velocity

David Lafferty[*]

*Institute for Theoretical Physics, Heidelberg University, Philosophenweg 16, 69120 Heidelberg, Germany*

Alexander Rothkopf[†]

*Faculty of Science and Technology, University of Stavanger, 4021 Stavanger, Norway*

We explore the in-medium properties of heavy-quarkonium states at finite baryochemical potential and finite transverse momentum based on a modern complex-valued potential model. Our starting point is a novel, rigorous derivation of the generalized Gauss law for in-medium quarkonium, combining the nonperturbative physics of the vacuum bound state with a weak coupling description of the medium degrees of freedom. Its relation to previous models in the literature is discussed. We show that our approach is able to reproduce the complex lattice QCD heavy quark potential even in the nonperturbative regime, using a single temperature dependent parameter, the Debye mass $m_D$. After vetting the Gauss law potential with state-of-the-art lattice QCD data, we extend it to the regime of finite baryon density and finite velocity, currently inaccessible to first principles simulations. In-medium spectral functions computed from the Gauss law potential are subsequently used to estimate the $\psi'/J/\psi$ ratio in heavy-ion collisions at different beam energies and transverse momenta. We find qualitative agreement with the predictions from the statistical model of hadronization for the $\sqrt{s_{NN}}$ dependence and a mild dependence on the transverse momentum.

## I. INTRODUCTION

Heavy quarkonium, the bound states of a charm or bottom quark, and its antiquark ($c\bar{c}$ or $b\bar{b}$) have matured into a high precision tool for the study of strongly interacting matter under extreme conditions in the context of relativistic heavy-ion collisions [1,2]. On the experimental side, the predominant decay of heavy quarkonia into dileptons makes them well-controlled observables, probing different stages of the quark gluon plasma (QGP) created in such collisions. The now iconic measurements of the dimuon spectrum of bottomonium by the CMS Collaboration [3] lend themselves to a phenomenological interpretation of the $b\bar{b}$ pair traversing the medium as a test particle (see for example [4]), sampling the whole history of the QCD medium. On the other hand, the more recent measurement of a finite elliptic flow of the $J/\psi$ particle by the ALICE Collaboration [5] indicates at least a partial equilibration of the charm quarks with their surrounding. The inevitable loss of memory about the initial conditions accompanying equilibration hence positions the lighter flavor as a probe of the late stages of the collision.

From a theory standpoint the heavy mass of the constituent quarks opens the door to the powerful effective field theory (EFT) framework [6] that allow us to simplify the description of their (non)equilibrium behavior. These techniques have led to progress both in direct lattice QCD studies of equilibrated quarkonium and in formulating real-time descriptions of their nonequilibrium evolution. The foundation of the EFT strategy is the presence of the natural separation of scales $m_Q \gg m_Q v \gg m_Q v^2$, with $m_Q$ the heavy quark mass and $v$ its typical velocity. These are denoted the hard (rest-energy), soft (momentum exchange), and ultrasoft (binding energy) scales, respectively. In addition to these three scales, there exists the characteristic scale of quantum fluctuations $\Lambda_{\rm QCD}$ and of thermal fluctuations $T$.

The construction of an EFT involves first choosing a cutoff energy above which the physics is not treated explicitly, before identifying the relevant degrees of freedom in the lower energy scale and treating these explicitly by writing down the most general Lagrangian compatible with the corresponding symmetries. Each term receives a complex-valued Wilson coefficient which needs to be determined by a matching procedure; that is, we compute a correlation function in the full microscopic theory and a correlator with the same physics content in the EFT and require that they agree below a certain scale. By this *integrating out* of higher scales, the results of the microscopic theory can be reproduced in the EFT as long as we stay within its range of validity at low energies.

[*]lafferty@thphys.uni-heidelberg.de
[†]alexander.rothkopf@uis.no

For heavy quarkonium this program has been implemented by integrating out the hard scale $\sim m_Q$ from the full QCD Lagrangian to give nonrelativistic QCD (NRQCD), a theory of nonrelativistic Pauli spinor fields. This can be achieved in a fully nonperturbative manner. In a further step, integrating out the soft scale $\sim m_Q v$ results in potential nonrelativistic QCD (pNRQCD) [7], where the potential governing the quarkonium dynamics becomes one of the Wilson coefficients determined via matching. While the perturbative derivation of pNRQCD has been successfully completed, its nonperturbative definition is still an active field of research. In this work we will utilize the fact that the in-medium potential between two static quarks can indeed be systematically derived from QCD in a nonperturbative fashion via the language of EFTs [8].

Much progress has been made in understanding the properties of equilibrated heavy quarkonium in a static medium from first principles by using lattice QCD computations. To directly study the in-medium modification of charmonium it is nowadays possible to deploy fully relativistic formulations of the heavy quarks (for some recent works see [11–13]). However, realistic simulations of bottomonium currently deploy lattice regularized versions of NRQCD [14,15]. In-medium correlation functions computed from first principles have revealed the presence of statistically significant in-medium modifications consistent with our intuition: the hotter the medium, the stronger the modification, and the more weakly bound the quarkonium state is in vacuum, the easier it is influenced by the medium.

Going beyond statements of overall in-medium modification remains difficult in lattice simulations, as it requires one to extract the spectral functions of in-medium quarkonium from Euclidean time correlators. This amounts to an exponentially hard ill-posed inverse problem, which despite concerted efforts of the community so far has only revealed insight into the properties of the ground state in-medium properties of different quarkonium channels. The most recent study of bottomonium and charmonium from lattice NRQCD has elucidated the change in mass of the ground state in-medium, concluding that quarkonium becomes lighter as temperature increases [15].

What all of these studies are still missing is access to the remnants of excited states in the medium, as well as to the threshold, which plays an important role in understanding the stability of the in-medium states. Indeed the in-medium binding energy is defined from the distance of the in-medium quarkonium spectral peak to the onset of the threshold. Currently this information is only accessible in potential based computations, where a Schrödinger equation for the spectral functions is solved in the presence of a nonperturbatively defined in-medium potential.

Important progress in this regard has been made using an EFT based definition of the in-medium potential between two static quarks based on the real-time evolution of the QCD Wilson loop,

$$V(r) = \lim_{t\to\infty} \frac{i\partial_t W_\square(r,t)}{W_\square(r,t)}. \quad (1)$$

Evaluating this expression in hard-thermal loop (HTL) resummed perturbation theory revealed [6,16,17] that in general the proper potential is a complex quantity with a real part exhibiting Debye screening and an imaginary part growing monotonously with temperature. The physical processes contributing to Im$V$ differ according to the separation of scales present [18]; the scattering of medium partons with the gluons mediating the heavy quark interaction, so-called Landau damping, as well as the gluon induced transition from a color singlet to octet may both contribute. At high temperature, Landau damping dominates and the potential reads

$$V_{\rm HTL}(r) = -\tilde{\alpha}_s\left[m_D + \frac{e^{-m_D r}}{r} + iT\phi(m_D r)\right] + \mathcal{O}(g^4), \quad (2)$$

where

$$\phi(x) = 2\int_0^\infty dz \frac{z}{(z^2+1)^2}\left(1 - \frac{\sin(xz)}{xz}\right). \quad (3)$$

Note that this potential describes the real-time evolution of an unequal time quarkonium correlation function and not of the wave function itself. Therefore the presence of an imaginary part is not directly related to the disappearance of the heavy quarks from the system but instead encodes the decoherence of the system from its initial state as it evolves over time in the thermal medium [19,20]. It is important to keep in mind that when we solve a Schrödinger equation with this potential, the resulting correlation function can be used to straightforwardly compute the in-medium quarkonium spectral function. The question of how to relate this complex-valued potential to the real-time evolution of the wave function is an active field of research, and recent progress has been made by considering the concept of open-quantum systems [21–25].

The real-time definition Eq. (1) is not directly amenable to an evaluation in lattice simulations since they are carried out in an artificial Euclidean (imaginary) time. Instead a strategy has been developed to use Bayesian inference to extract the values of the potential nonperturbatively on the lattice [26–28]. Many studies have confirmed (see for example [29]) that at low temperatures the potential is well described by the Cornell form

$$V^{\text{vac}}(r) = -\frac{\tilde{\alpha}_s}{r} + \sigma r + c, \quad (4)$$

where $\bar{\alpha}_s = C_F g^2/4\pi$ is the strong coupling constant (the factor $C_F$ has been absorbed to match the phenomenology literature), $\sigma$ is the string tension, and $c$ is an additive constant that will be used for calibration purposes. Equation (4) already captures the two most important features of QCD: asymptotic freedom via the running coupling at small distances and confinement via the nonperturbative linear rise.

At finite temperature lattice QCD tells us that not only the real part weakens gradually as one moves into the deconfined phase but that above the pseudocritical temperature indeed a finite imaginary part is present [29,30]. To put these numerical results to use in computations of quarkonium spectral functions an efficient analytic parametrization of the complex valued potential is needed. Deploying it in a Schrödinger equation gives access to spectral functions from which we may learn about physically relevant properties, such as in-medium masses or decay widths.

To this end, in [31] one of the authors proposed a simple model of a nonperturbative vacuum bound state immersed in a weakly coupled medium (see also [32,33]). The former is described by the Cornell potential, the latter by an in-medium permittivity evaluated in HTL perturbation theory. In that model the effects of the medium on the vacuum potential are incorporated by the application of the generalized Gauss law [34]. Once the vacuum parameters of the Cornell potential are chosen, the model provides a prediction for the full in-medium values of Re$V$ and Im$V$ based on a single temperature dependent parameter, identified as the Debye mass $m_D$. While it is not obvious that such an ansatz can accommodate the physics of the heavy quark potential, especially in the nonperturbative regime close to the crossover temperature, it has been shown that tuning of $m_D$ reproduces the lattice QCD values of Re$V$ and the tentative values of Im$V$ quite well. In turn the Gauss law model has been used to study the in-medium properties of quarkonium in a thermal medium, as well as to provide an estimate for the $\psi'/J/\psi$ ratio at very high energy heavy-ion collisions at midrapidity and zero transverse momentum.

However, there exists two shortcomings of the Gauss law model of [31], one technical and one phenomenological, which limit its utility in exploring heavy quarkonium in heavy-ion collisions. The technical one is related to the fact that in order to derive the in-medium modification of the string part of the Cornell potential, the previous study introduced an *ad hoc* assumption about the functional form of the in-medium permittivity in coordinate space. On the phenomenological side, the model did not incorporate the effects of finite baryochemical potential or transverse momentum, both of relevance to compare to actual data from heavy-ion collisions.

The present study sets out to overcome both of these issues. As a first step we put forward a novel and improved Gauss law model, based on a more rigorous derivation, which does not rely on any *ad hoc* ingredients. Taking into account string breaking, i.e., the fact that the vacuum Cornell potential does not rise indefinitely, we are able to derive well-defined expressions for Re$V$ and Im$V$. Their functional form turns out to be simpler than the one obtained in [31]. Again, Re$V$ and Im$V$ only depend on a single temperature dependent parameter $m_D$. They furthermore consistently reduce to the HTL result at large values of the Debye mass parameter (high temperature) and to the Cornell potential at vanishing $m_D$ (vacuum). Using state-of-the-art lattice QCD results for Re$V$ and Im$V$ we show that the new model reproduces the nonperturbative values excellently by an appropriate selection of $m_D$. The relation of this new Gauss law model to previous model potentials in the literature is also discussed.

The second improvement is related to extending the model to settings not accessible to first principle lattice QCD simulations, i.e., quarkonium in a medium at finite baryochemical potential, as well as quarkonium traversing the QGP with a finite velocity. The latter is implemented via the hot-wind scenario, where the medium moves with a finite relative velocity with respect to the quarkonium [35]. The extended model will be used to compute the in-medium spectral functions of both charmonium and bottomonium states and investigate their in-medium properties as well as their melting. Modeling the finite baryochemical potential regime of the QCD phase diagram allows us to estimate the $\psi'/J/\psi$ ratio in heavy-ion collisions at lower beam energies, relevant for future collider facilities, such as FAIR and NICA. By modeling quarkonium at finite velocity we give predictions for the $\psi'/J/\psi$ ratio.

## II. THE GAUSS LAW POTENTIAL MODEL

In order to fully utilize the advances in lattice QCD computations of the in-medium heavy quark potential in phenomenological studies, an analytic parametrization of the complex $V(r)$ is required. Such a parametrization may also provide a starting point for modeling heavy quark interactions in regions of the QCD phase diagram currently inaccessible to first principles Monte Carlo simulations.

More specifically, we require easily evaluable expressions for both the real and the imaginary parts of the in-medium heavy-quark potential that provide a faithful reproduction of nonperturbative lattice data where available. In particular it must be applicable in the temperature regime close to and around the crossover transition, where HTL perturbation theory by itself is not valid. In this study we deploy a similar physical reasoning as in [29,31] to construct such an analytic parametrization of the potential, overcoming the previous shortcoming in that we avoid any *ad hoc* assumptions on the linear part of the in-medium potential.

The starting point is the fact that the vacuum behavior of quarkonium bound states is described well by the Cornell potential of Eq. (4). We then consider this heavy-quark two-body system immersed in a weakly coupled medium described by the HTL permittivity. Both the Coulombic and the stringlike part of the Cornell potential will then receive in-medium corrections, which we compute using linear response theory.

### A. Constructing the in-medium model

In linear response theory the electric field at finite temperature, or equivalently the electric potential, can be obtained from its vacuum counterpart by multiplying it by the inverse of the static dielectric constant in momentum space [36],

$$V(\mathbf{p}) = \frac{V^{\text{vac}}(\mathbf{p})}{\varepsilon(\mathbf{p}, m_D)}. \quad (5)$$

The permittivity, defined as an appropriate limit of the real-time in-medium gluon propagator, imprints the medium effects onto the potential here. In the following we will consider separately the Coulomb and stringlike parts of the Cornell potential in the above relation.

Note that Eq. (5) does not rely on a weak-coupling assumption and remains valid as long as the vacuum field is weak enough to justify the linear approximation. Using the convolution theorem it can be recast in coordinate space as follows:

$$V(\mathbf{r}) = (V^{\text{vac}} * \varepsilon^{-1})(\mathbf{r}), \quad (6)$$

where "$*$" represents the convolution.

To continue we consider the other central building block of our approach, the generalized Gauss law,

$$\nabla \cdot \left(\frac{\mathbf{E}^{\text{vac}}}{r^{a+1}}\right) = 4\pi q \delta(\mathbf{r}), \quad (7)$$

which holds for electric fields of the form $\mathbf{E}^{\text{vac}}(r) = -\nabla V^{\text{vac}}(r) = qr^{a-1}\hat{r}$. This reduces to the well-known form for Coulombic potentials with $a = -1$, $q = \tilde{\alpha}_s$), while the linearly rising string case corresponds to $(a = 1, q = \sigma)$. For a general $a$ we have

$$-\frac{1}{r^{a+1}}\nabla^2 V^{\text{vac}}(r) + \frac{1+a}{r^{a+2}}\nabla V^{\text{vac}}(r) = 4\pi q \delta(\mathbf{r}). \quad (8)$$

Denoting the differential operator on the left-hand side above as $\mathcal{G}_a$, we now apply it to Eq. (6) to deduce the general integral expressions for each term in the in-medium heavy-quark potential,

$$\mathcal{G}_a[V(r)] = \mathcal{G}_a \int d^3y (V^{\text{vac}}(r-y)\varepsilon^{-1}(y))$$
$$= 4\pi q(\delta * \varepsilon^{-1})(r)$$
$$= 4\pi q \varepsilon^{-1}(r, m_D), \quad (9)$$

where we have used Eq. (8) and that the convolution commutes with $\mathcal{G}_a$. For the Coulombic and string cases, respectively, this gives

$$-\nabla^2 V_C(r) = 4\pi\tilde{\alpha}_s \varepsilon^{-1}(r, m_D), \quad (10)$$

$$-\frac{1}{r^2}\frac{d^2 V_S(r)}{dr^2} = 4\pi\sigma\varepsilon^{-1}(r, m_D). \quad (11)$$

At this point we introduce the explicit expression for the coordinate space in-medium permittivity obtained from the perturbative HTL expression in momentum space [36],

$$\varepsilon^{-1}(p, m_D) = \frac{p^2}{p^2 + m_D^2} - i\pi T \frac{pm_D^2}{(p^2 + m_D^2)^2}. \quad (12)$$

The real part of the real space expression can be calculated via a contour integral and the residue theorem to give

$$\text{Re}\varepsilon^{-1}(r, m_D) = -\frac{m_D^2 e^{-m_D r}}{4\pi r}, \quad (13)$$

while the imaginary part is expressed by a Meijer-$G$ function,

$$\text{Im}\varepsilon^{-1}(r, m_D) = -\frac{m_D T}{4r\sqrt{\pi}} G_{1,3}^{2,1}\left(\begin{array}{c} -\frac{1}{2} \\ -\frac{1}{2}, -\frac{1}{2}, 0 \end{array} \middle| \frac{1}{4} m_D^2 r^2\right). \quad (14)$$

Let us use Eqs. (13) and (14) to solve for the in-medium modified Coulombic part of the potential. We find that our ansatz, as expected, reproduces the well-known HTL result [17,37] given in Eq. (2),

$$\text{Re}V_C(r) = -\tilde{\alpha}_s\left[m_D + \frac{e^{-m_D r}}{r}\right], \quad (15)$$

$$\text{Im}V_C(r) = -\tilde{\alpha}_s[iT\phi(m_D r)], \quad (16)$$

where $\phi$ is defined in Eq. (3). The next step is to turn to the string part, Eq. (11), for which the formal solution can be straightforwardly written down as

$$V_S(r) = c_0 + c_1 r - 4\pi\sigma \int_0^r dr' \int_0^{r'} dr'' r''^2 \varepsilon^{-1}(r'', m_D). \quad (17)$$

The constants $c_0$ and $c_1$ will be chosen to ensure the physically motivated boundary conditions $\text{Re}V_S(r)|_{r=0} = 0$, $\text{Im}V_S(r)|_{r=0} = 0$, and $\partial_r \text{Im}V_S(r)|_{r=0} = 0$. This leads to the following analytical form:

$$\mathrm{Re}V_S(r) = \frac{2\sigma}{m_D} - \frac{e^{-m_D r}(2 + m_D r)\sigma}{m_D}, \quad (18)$$

$$\mathrm{Im}V_S(r) = \frac{\sqrt{\pi}}{4} m_D T \sigma r^3 G_{2,4}^{2,2}\left(\begin{array}{c}-\frac{1}{2},-\frac{1}{2}\\ \frac{1}{2},\frac{1}{2},-\frac{3}{2},-1\end{array}\bigg|\frac{1}{4}m_D^2 r^2\right). \quad (19)$$

With both the Cornell and the string in-medium solutions at hand we combine the expressions to form the full Gauss law model

$$\mathrm{Re}V = \mathrm{Re}V_C + \mathrm{Re}V_S + c, \quad \mathrm{Im}V = \mathrm{Im}V_C + \mathrm{Im}V_S. \quad (20)$$

Let us inspect the properties of each contribution to the in-medium potential found so far. For the real parts, the short distance $r \to 0$ limit, corresponding to each heavy-quark not being able to "see" the intermediate medium, recovers the vacuum Cornell potential. The zero temperature limit corresponds to $m_D \to 0$, which also recovers the Cornell potential. At large distances the real part exhibits an exponential flattening off $\sim e^{-m_D r}$, which is the characteristic and well-known Debye screening behavior. Due to the extra factor of $m_D$ in the denominator, the string contribution to the in-medium real-part will become more and more suppressed at high temperature, eventually giving way to the pure HTL result. The imaginary part arising from the Coulombic contribution asymptotes to a constant at large distances, which is expected for Landau damping. Only the string imaginary part, Eq. (19), at first sight appears problematic, as it diverges logarithmically at large $r$. As we argue in the next section this is a manifestation of the absence of string breaking in the Cornell potential, and we will account for it by introducing a well-motivated regularization.

### B. Consistent treatment of string breaking

In the preceding section we found that only the string imaginary part, Eq. (19), shows an unphysical behavior, in that it diverges logarithmically at large $r$. To understand the origin of this artifact let us consider the ingredients used. The generalized Gauss law [Eq. (7)] is formally correct, and the linear-response relation in Eq. (5) is valid under a weak-field ansatz. However, the vacuum potential and in-medium permittivity both operate under assumptions that can be challenged. First, we have utilized a vacuum potential that contains an unending and unphysical linear rise. Second, the expression for the complex permittivity given in Eq. (12) is a hard-thermal-loop result and as such is strictly only valid at temperatures much larger than $T_c$. We contend that the combination of these two assumptions leads to the unwanted infrared divergence in the final expression for the string imaginary part.

In our computation, both issues manifest themselves in Eq. (17), which can be written, after substituting the imaginary part of Eq. (12) into Eq. (17) and performing the angular integration of the inverse Fourier transform, as follows:

$$\mathrm{Im}V_S(r) = c_0 + c_1 r + 2T\sigma m_D^2 \int_0^r \mathrm{d}r' \int_0^{r'} \mathrm{d}r'' r''^2 \\ \times \int_0^\infty \mathrm{d}p\, p^2 \frac{\sin(pr'')}{pr''} p^2 \frac{1}{p(p^2 + m_D^2)^2}. \quad (21)$$

We have arranged the momentum factors as above to make clear their different origins: the first term ($p^2$) arises from integrating in spherical coordinates and the second $[\sin(pr'')/pr'']$ after completing the polar integration. The last two terms encapsulate the contribution from the gluon propagator, and it is the $1/p$ factor here that we identify as causing the weak infrared divergence. In order to regularize this integral, we modify the last term as follows:

$$\frac{1}{p(p^2 + m_D^2)^2} \to \frac{1}{\sqrt{p^2 + \Delta^2}(p^2 + m_D^2)^2}, \quad (22)$$

where ($\Delta$) will be a suitably chosen regularization scale. In Eq. (21) the spatial integrals can be carried out analytically, which combined with the regularization above gives our new definition of the string imaginary part,

$$\mathrm{Im}V_S(r, \Delta) = 2T\sigma m_D^2 \int_0^\infty \mathrm{d}p\, \frac{2 - 2\cos(pr) - pr\sin(pr)}{\sqrt{p^2 + \Delta^2}(p^2 + m_D^2)^2}, \quad (23)$$

where we have also chosen the constant terms to impose the boundary conditions as before. Equation (23) can be numerically evaluated very quickly, and the only remaining step is now to choose the regularization scale $\Delta$.

To this end we propose the following physically motivated scheme. Note that if we rescale momentum via $p \to p/m_D$ and rearrange slightly, Eq. (23) takes on a suggestive form:

$$\mathrm{Im}V_S(r, \Delta_D) = \frac{\sigma T}{m_D^2} \chi(m_D r, \Delta_D), \quad (24)$$

where

$$\chi(x) = 2 \int_0^\infty \mathrm{d}p\, \frac{2 - 2\cos(px) - px\sin(px)}{\sqrt{p^2 + \Delta_D^2}(p^2 + 1)^2} \quad (25)$$

and $\Delta_D = \Delta/m_D$. That is, we can express the string imaginary part as some temperature dependent prefactor with dimensions of energy, multiplied by a dimensionless momentum integral. This is identical to the Coulombic expression, where the integral asymptotes to unity in the limit $r \to \infty$. We thus impose the same condition for the string part.

This procedure also recovers the correct behavior at large $T$ (large $m_D$); i.e., the string contribution to the imaginary part diminishes while the Coulombic part grows in stature and we eventually recover the pure HTL result. The value of the regularization parameter $\Delta_D$ can be computed numerically (the trigonometric terms in Eq. (25) drop out in the $r \to \infty$ limit). Furthermore, since it is expressed in terms of the Debye mass, it remains constant and the computation need be performed only once. We find

$$\Delta_D = \Delta/m_D \simeq 3.0369. \quad (26)$$

In order to check whether our particular choice of regularization influences the end result we also implemented instead a factor of $\tanh(p/\Delta')$ in the integral in Eq. (21). The hyperbolic tangent rises linearly at small $p$ and converges exponentially quickly to unity and thus is able to fix the infrared divergence while leaving the ultraviolet behavior unchanged. We find that using this alternative leads to an equivalent $\Delta'_D$ that lies within 1% of the value in Eq. (26), and the subsequent results for Im$V$ are indistinguishable by eye to those obtained via the original method. That the regularization process is independent of the exact technique used serves as an encouraging cross-check.

Figure 1 depicts the values of the real (left column) and imaginary (right column) part of our novel Gauss law parametrization at three different realistic combinations of temperature and Debye mass. The red solid lines correspond to the Coulombic contributions and the green lines to the string parts. The total is shown as a black solid line. In the panels for Re$V$ the vacuum Cornell potential is added as a gray solid line. For completeness the unregularized string imaginary part is included as a blue line. The figures clearly show how the perturbative HTL results, i.e., the purely Coulombic contribution, dominate at high temperatures.

### C. Comparison to other models

Let us compare the expressions derived above from the Gauss law with other models previously deployed in the study of heavy quarkonium. Before the realization that the in-medium potential is a complex quantity, the focus lay on modeling a real-valued potential. A classic work in this regard is the study by Karsch, Mehr, and Satz (KMS) [38], who argued based on the two-dimensional Schwinger model that the in-medium potential should show both the standard Debye screening term for the Coulombic term and an exponential damping factor for the string part:

$$V^{\text{KMS}}(r) = \frac{\sigma}{m_D}(1 - e^{-m_D r}) - \frac{\tilde{\alpha}_s}{r} e^{-m_D r}. \quad (27)$$

Equation (27) reduces (by construction) to the Cornell potential in the limit $m_D \to 0$. At first sight the exponential damping of the string part may appear similar to our result for $V_S(r)$. A quantitative inspection, however, shows that the KMS potential exhibits a stronger dependence on $m_D$; i.e., for the same value of $m_D$ the deviation from the $m_D = 0$ limit is more pronounced than in our case.

The KMS potential may also be obtained, as presented in version 1 of [39], by modeling the interquark interactions as an effective one-dimensional stringlike interaction. In addition, the authors postulate that entropy may contribute to the interquark interaction. Their heuristic arguments, resting on an identification of the real part of the potential with thermodynamic quantities, leads them to propose for the in-medium string part

$$\text{Re}V^{\text{GPDM}}(r) = \frac{2\sigma}{m_D}(1 - e^{-m_D r}) - \sigma r e^{-m_D r}. \quad (28)$$

This expression turns out to be the same as our result, which in this paper has been systematically derived from the Gauss law ansatz. We are thus able to offer an explanation for the presence of the additional $r$ dependent term that does not rely on the *ad hoc* identification of the real part of the potential with either the free or internal energies of the quarkonium system.

In the context of purely real-valued in-medium potential models, the generalized Gauss law was used for the first time in [32]. In this study a first attempt was made to combine the Gauss law and Debye-Hückel theory to implement the screening of both the Coulombic and stringlike parts of the potential. This represented an important step forward toward a systematic modeling of both terms at finite temperature. The authors encountered difficulty in using this parametrization to capture the behavior of the color singlet free energy in lattice QCD, which was taken as a proxy for the in-medium real part of the potential. This led to the introduction of an additional parameter $\kappa$ to compensate for these deviations. Unfortunately at that time it was not possible to relate $\kappa$ either to the parameters of the Cornell potential or to the Debye mass.

An important step toward parametrizing the interquark potential as a complex quantity was taken in [33], where the authors proposed taking the linear response relation in Eq. (5) at face value and introduced the same HTL permittivity as used in this paper to directly carry out the inverse Fourier transform. Their computation led to the following proposal for the in-medium potential:

$$\text{Re}V(r) = -\tilde{\alpha}_s m_D \left( \frac{e^{-m_D r}}{r} + 1 \right) \quad (29)$$

$$+ \frac{2\sigma}{m_D}\left(\frac{e^{-m_D r} - 1}{r} + 1\right), \quad (30)$$

$$\text{Im}V(r) = -\tilde{\alpha}_s T \phi(m_D r) + \frac{2\sigma T}{m_D^2}\chi_0(m_D r), \quad (31)$$

where $\phi$ was defined in Eq. (3) and $\chi_0$ is given by

$$\chi_0(x) = 2\int_0^\infty \frac{dz}{z(z^2+1)^2}\left(1 - \frac{\sin(xz)}{xz}\right). \quad (32)$$

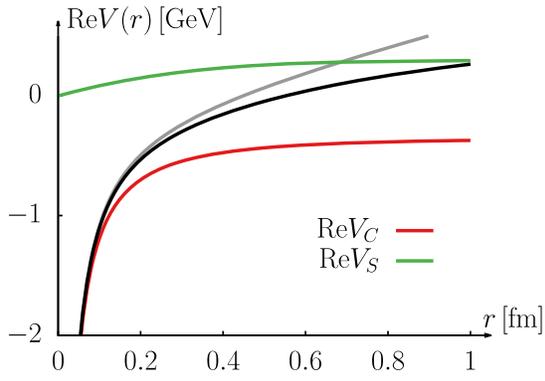
(a) $T = 200$ MeV, $m_D \simeq 380$ MeV

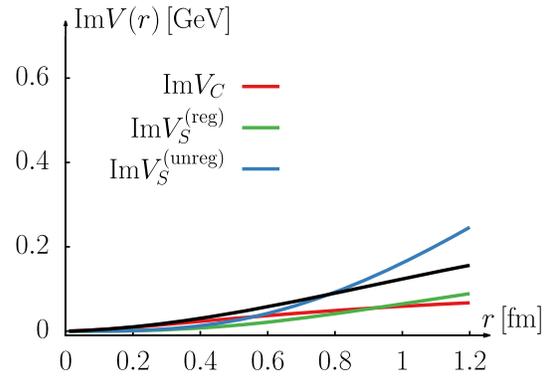
(b) $T = 200$ MeV, $m_D \simeq 380$ MeV

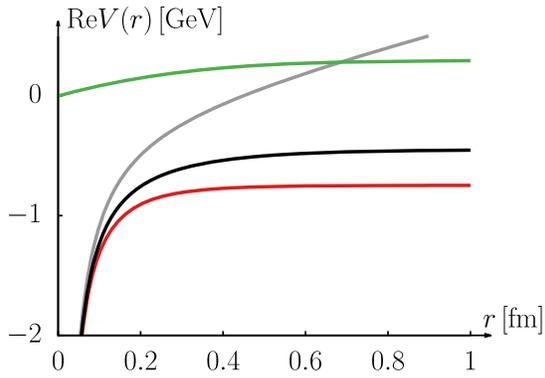
(c) $T = 500$ MeV, $m_D \simeq 1.15$ GeV

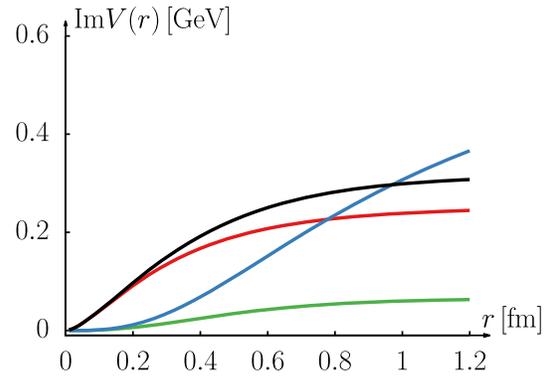
(d) $T = 500$ MeV, $m_D \simeq 1.15$ GeV

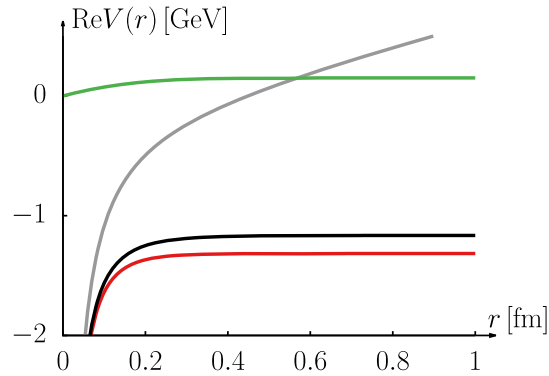
(e) $T = 1$ GeV, $m_D \simeq 2.25$ GeV

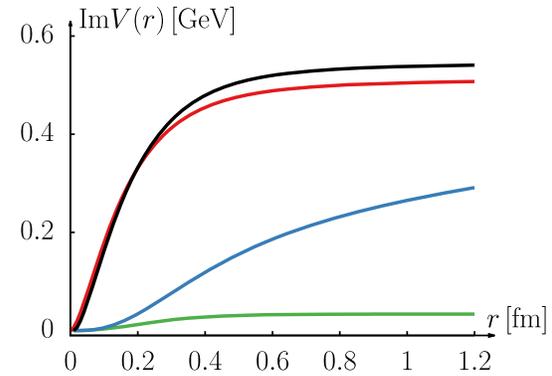
(f) $T = 1$ GeV, $m_D \simeq 2.25$ GeV

FIG. 1. Temperature dependence of the Coulombic (red curves), string regularized (green curves), and string unregularized (blue curves) imaginary parts. In the high temperature limit we recover the purely Coulombic HTL result.

In this approach the modification of the real and imaginary parts are governed by a single temperature dependent parameter, just as in our study.

If we inspect the functional form of this parametrization closer, two properties become apparent which challenge the validity of the result. One can be remedied but the other hints at a more foundational difficulty of the approach.

Let us first consider the imaginary part. Just as in our derivation, the string imaginary part of the potential diverges logarithmically due to the $1/z$ term. This divergence may be avoided, as discussed in the previous chapter, by regularizing the unphysical linear rise to infinity of the Cornell potential, i.e., by introducing string breaking. On the other hand, if we take a look at the in-medium real part arising from the vacuum stringlike potential, we find that it contains an unscreened $1/r$ term. This would suggest that the deconfined color charges are unable to screen the interactions and a long-range component remains in Re$V$. Such a behavior is counterintuitive and does not agree with current lattice data determinations of the in-medium potential.

The authors of [39] recently proposed a very different derivation of the real part in Eq. (28). They base it on the direct Fourier transform of the gluon propagator, which, however, receives an additional nonperturbative contribution originally suggested in [40]. The additional term in the gluon propagator is related to a nonvanishing gluon condensate, which has previously been used to justify a a similar real part deployed in the $T$-matrix approach in [10]. In addition that study models quarkonium screening with different screening masses for the Coulombic $m_D$ and stringlike part $m'_D$ and a third parameter $c_s$ to take into account string breaking effects,

$$\mathrm{Re}V^{\mathrm{RL}} = -\tilde{\alpha}_s \frac{e^{-m_D r}}{r} - \frac{\sigma}{m'_D} e^{-m'_D r - (c_s m'_D r)^2}. \quad (33)$$

In [31] the Gauss law was used for the first time in the context of a complex valued in-medium potential. The study brought together the ideas of the Debye-Hückel theory from [32] with the HTL permittivity as used in [33]. However, as already mentioned in the introductory section, to solve the Gauss law equations in that paper, the authors introduced *ad hoc* assumptions about the string part of the in-medium potential. The present paper, while using a very similar combination of HTL permittivity and linear response theory, provides a rigorous derivation of the in-medium expressions for Re$V$ [Eqs. (15) and (18)] and Im$V$ [Eqs. (16) and (23)] that forms the central result of this chapter.

### D. Vetting with lattice QCD data

The most important benchmark for any parametrization of the in-medium heavy quark potential is whether it is able to reproduce the nonperturbative lattice QCD results. Since our Gauss law approach uses the HTL permittivity to modify the nonperturbative vacuum potential, it is by no means obvious that it can capture the physics of the interquark potential in the nonperturbative regime around the crossover transition. We will show in this section that it indeed works excellently even in this regime.

One hint at why the Gauss law may work where HTL alone is no longer valid is given by the form of the Gauss law equation of the Coulombic part in coordinate space. As was discussed in [31], using the HTL permittivity leads to an expression that has the same form as a linear response equation for $V_C$ as the original Debye Hückel theory, and the Debye mass parameter governs the strength of the linear response. This allows one to smoothly connect the expression at finite $T$ to the unscreened potential at $T = 0$.

TABLE I. Best fit result for the vacuum potential parameters after fitting to the low temperature lattice QCD data.

|                | $\beta_1 = 6.9$ | $\beta_2 = 7.48$ |
| --- | --- | --- |
| $\tilde{\alpha}_s$ | $0.471 \pm 0.047$ | $0.385 \pm 0.027$ |
| $\sqrt{\sigma}$ [GeV] | $0.466 \pm 0.017$ | $0.515 \pm 0.014$ |
| $c$ [GeV] | $1.781 \pm 0.059$ | $2.648 \pm 0.042$ |

The vetting is carried out using published state-of-the-art lattice QCD data for the real part of the potential [29,30]. The values for Re$V$ and Im$V$ have been extracted from simulations of the HotQCD Collaboration on $48^3 \times 12$ lattices featuring $N_f = 2 + 1$ flavors of dynamical light quarks discretized with the asqtad [41] action. The pion mass on these lattices is larger than physical at $m_\pi \approx 300$ MeV, with a transition temperature $T_c \approx 175$ MeV. As the temperature is changed on these lattices by changing the lattice spacing, there are zero temperature ensembles available for calibration purposes.

The first step in applying the Gauss law parametrization consists of fixing the vacuum parameters $\tilde{\alpha}_s$, $\sigma$, and $c$ appearing in the Cornell potential, which characterize the test charges inserted into the medium. Two sets of low temperature results for Re$V$ are available [29], to which Eq. (4) is fitted. The results are given in Table I and shown in gray in Fig. 2. The naive Cornell ansatz works excellently in describing the lattice data in the phenomenologically relevant range of distances 0.1 fm $< r <$ 1.2 fm.

Once the vacuum parameters are determined, the Gauss law parametrization contains a single temperature dependent parameter, the Debye mass $m_D$, which controls both Im$V$ and Re$V$. Here we will fit the values of $m_D$ using the real part. As can be seen from the left panel of Fig. 2 the fit works excellently and the Gauss law parametrization is able to capture the behavior of the potential from the Coulombic region, through the intermediate regime and up to the screening regime at large distances and high temperatures. This gives us confidence that the derived expressions for the in-medium potential do capture the relevant physics encoded nonperturbatively in the lattice data. The best fit values for the Debye mass parameter are listed in Table II.

Once $m_D$ is fixed we can compare the corresponding Gauss law prediction for the imaginary part with the tentative values extracted on the lattice. As shown in the left panel of Fig. 2 we find that the agreement is also excellent down to $T = 164$ MeV, which on these lattices corresponds already to the confined phase. Only at $T = 148$ MeV, deep in the hadronic regime, do deviations start to appear. Note that the imaginary part in the Gauss law rises more steeply as the temperature increases but the asymptotic large distance value behaves nonmonotonously with the temperature, reflecting the competition between Im$V$ arising from the string and Coulombic parts of the potential.

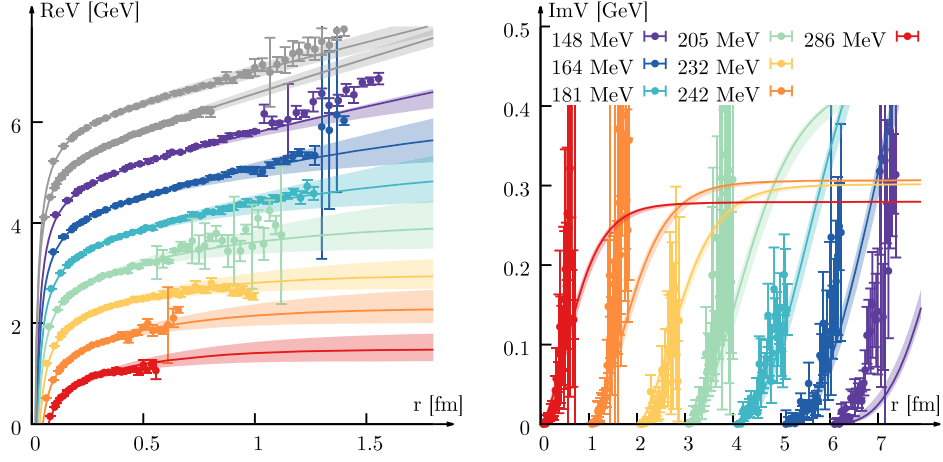

FIG. 2. (Left) The real part of the Gauss law model fitted to lattice QCD results. The three vacuum parameters are determined from $T = 0$ lattice data (gray). The finite temperature lattice data (colored points) are reproduced by tuning the $m_D$ parameter. Solid lines give the best fit results and the shaded regions the corresponding errors that arise from uncertainty both in the initial lattice data and in our vacuum parameters. (Right) Prediction of the in-medium imaginary part from the Gauss law model (solid lines) fixed by the values of $m_D$ obtained from Re$V$. Tentative lattice QCD results for Im$V$ show excellent agreement.

We have compared the best fit of the properly derived Gauss law expression to that obtained with the legacy formulation of [29]. Within the combined statistical and systematic errors, both satisfactorily reproduce the lattice data. That is, the uncertainty in the available values of Re$V$ does not yet allow us to favor one over the other. We note that the two best fit solutions start to deviate from each other for $r \gtrsim 0.6$ fm (the QGP phase), leading to differences in their asymptotic values. This in turn translates into quantitative differences in the precise temperature dependence of the open-heavy flavor threshold and thus the binding energy of the in-medium quarkonium states. It will require future high precision lattice determinations of Re$V$ to distances up to $r \sim 1$ fm) to resolve this phenomenologically relevant ambiguity.

### E. Extension to a running coupling

In anticipation of upcoming high resolution lattice QCD computations of the in-medium heavy quark potential, it is prudent to consider the effects of a running coupling in the Gauss law parametrization. While in the simulation data deployed in the previous section the short distance regime was still well described by a naive Cornell potential, more recent lattice studies of heavy quark interactions [42] have shown that at shorter resolved distances the running will manifest itself. Thus we consider the strong coupling parameter of our Cornell potential to become a function of distance $\tilde{\alpha}_s \to \tilde{\alpha}_s(r)$ and write

$$\tilde{\alpha}_s(r) = \cdots + \frac{\tilde{\alpha}_s^{(-1)}}{r} + \tilde{\alpha}_s^{(0)} + \tilde{\alpha}_s^{(1)} r + \tilde{\alpha}_s^{(2)} r^2 + \cdots. \quad (34)$$

Note that in the context of the vacuum potential in Eq. (4), we have already implicitly included the terms $\tilde{\alpha}_s^{(1)}$ and $\tilde{\alpha}_s^{(2)}$ by absorbing them into the other vacuum parameters.

In a thermal setting, this would necessitate including $r^a$ terms other than $a = -1, 1$ in the formulation of the in-medium potential. To do this, we must use the generalized Gauss law operator $\mathcal{G}_a$ given in the left-hand side of Eq. (8), but with a modified right-hand side that includes the real-space complex permittivity (following the procedure in Sec. II A)

$$-\frac{1}{r^{a+1}} \nabla^2 V(r) + \frac{1+a}{r^{a+2}} \nabla V(r) = 4\pi q \varepsilon^{-1}(r, m_D). \quad (35)$$

With the real space expressions given in Eqs. (13) and (14), a computer algebra program will give a general solution for general $a$ as follows:

$$\text{Re}V_a(r) = c_0 + c_a \frac{r^a}{a}$$
$$- \frac{q}{(m_D)^a}[\Gamma(a, m_D r) + \Gamma(1 + a, m_D r)], \quad (36)$$

TABLE II. Results for the in-medium potential parameters.

| $\beta$ | 6.8 | 6.9 | 7 | 7.125 | 7.25 | 7.3 | 7.48 |
|---|---|---|---|---|---|---|---|
| $T/T_c$ | 0.86 | 0.95 | 1.06 | 1.19 | 1.34 | 1.41 | 1.66 |
| $m_D/\sqrt{\sigma}$ | 0.153(13) | 0.403(33) | 0.537(42) | 0.769(56) | 1.062(72) | 1.081(72) | 1.297(79) |
| $m_D/T$ | 0.473 | 1.143 | 1.401 | 1.818 | 2.273 | 2.229 | 2.334 |

$$\mathrm{Im}V_a(r) = c_0 + \frac{1}{m_D}\left[c_a \frac{m_D r^a}{a} - \sqrt{\pi}qr^a TG^{2,2}_{2,4}\left(\begin{matrix}\frac{1}{2}, 1-\frac{a}{2} \\ \frac{3}{2}, \frac{3}{2}, 0, -\frac{a}{2}\end{matrix}\middle| \frac{1}{4}m_D^2 r^2\right)\right], \quad (37)$$

where $G$ is again the Meijer-$G$ function and $\Gamma$ is the upper incomplete Gamma function defined via

$$\Gamma(s,x) = \int_x^\infty \mathrm{d}t\, t^{s-1}e^{-t}. \quad (38)$$

Note that the imaginary part of the HTL solution expressed via the integral Eq. (3) is equivalent to Eq. (37) with ($a = 1$), as noted by the authors in [17].

Two remarks are in order: first, there is nothing in principle that prohibits the expressions above from being used throughout the remaining analysis in this chapter. We have decided not to do so because currently published lattice data on the in-medium potential does not yet reach the regime where the running is significant. Second, for $a \leq 1$ the imaginary part exhibits a divergence. In order to employ Eq. (37) in a phenomenological study one would need to carry out a regularization procedure similar to that discussed in the last section, which is in principle possible, but we have not investigated this further.

Keeping in mind that the Gauss law approach can straightforwardly be extended to accommodate a running coupling we nevertheless proceed to use its naive formulation ($a = -1, 1$ only) to investigate the in-medium properties of heavy quarkonium in subsequent chapters.

### F. Extension to finite velocity

In preparation for the study of heavy quarkonium at finite transverse momentum we need to consider how to extend the Gauss law model to treat a heavy quarkonium bound state moving through the plasma at finite velocity. To this end we will follow the ideas laid out in [43]. That is, one considers a QCD plasma in thermal equilibrium and a reference frame in which the medium moves at velocity **v** with respect to the bound state at rest, a so-called hot wind scenario. It then becomes necessary to distinguish two separate alignments: one in which the medium velocity is parallel to the dipole axis of the bound state and another in which it is perpendicular. This leads to distinct self-energies with different angular dependencies and, correspondingly, different expressions for the potential in each of the alignments. As reviewed in detail in Appendix the corresponding in-medium permittivity can be computed for both alignments and used to set up a finite-velocity Gauss law model in the same manner as in Sec. II A.

For the parallel alignment case, we find for the real and imaginary parts of the Coulombic part, respectively:

$$\mathrm{Re}V_C(\mathbf{r} \parallel \mathbf{v}) = c - \tilde{\alpha}_s m_D - \frac{\tilde{\alpha}_s}{r} + \tilde{\alpha}_s \int_0^{\pi/2} \mathrm{d}\theta\, \sin(\theta)\mathrm{Re}\left[\sqrt{\Pi_R^\parallel(\theta, v)}\right] e^{-\mathrm{Re}\left[\sqrt{\Pi_R^\parallel(\theta, v)}\right] r\cos(\theta)}, \quad (39)$$

$$\mathrm{Im}V_C(\mathbf{r} \parallel \mathbf{v}) = 2\tilde{\alpha}_s T \int_0^{\pi/2} \mathrm{d}\theta\, \sin(\theta)\frac{(1-v^2)^{3/2}(2+v^2\sin^2(\theta))}{2(1-v^2\sin^2(\theta))^{5/2}}\frac{m_D^2}{2i\mathrm{Im}\Pi_R^\parallel(\theta, v)}$$
$$\times \left\{\left[\sinh\left(r\cos(\theta)\sqrt{\Pi_R^\parallel(\theta, v)}\right)\mathrm{Shi}\left(r\cos(\theta)\sqrt{\Pi_R^\parallel(\theta, v)}\right)\right.\right.$$
$$-\cosh\left(r\cos(\theta)\sqrt{\Pi_R^\parallel(\theta, v)}\right)\mathrm{Chi}\left(r\cos(\theta)\sqrt{\Pi_R^\parallel(\theta, v)}\right)\bigg]$$
$$-\left[\sinh\left(r\cos(\theta)\sqrt{\Pi_R^\parallel(\theta, v)^*}\right)\mathrm{Shi}\left(r\cos(\theta)\sqrt{\Pi_R^\parallel(\theta, v)^*}\right)\right.$$
$$\left.\left.+\cosh\left(r\cos(\theta)\sqrt{\Pi_R^\parallel(\theta, v)^*}\right)\mathrm{Chi}\left(r\cos(\theta)\sqrt{\Pi_R^\parallel(\theta, v)^*}\right)\right]\right\}. \quad (40)$$

The retarded self-energy $\Pi_R^\parallel$ is given in Eq. (A14) and the functions $\mathrm{Shi}(x)$ and $\mathrm{Chi}(x)$ are defined in Eqs. (A30) and (A31). For the perpendicular case the results are identical after the replacements

$$\int_0^{\pi/2} \mathrm{d}\theta \to \int_0^{\pi/2} \mathrm{d}\theta \int_0^{2\pi} \frac{\mathrm{d}\phi}{2\pi}, \quad (41)$$

$$\Pi_R^\parallel(\theta, v) \to \Pi_R^\perp(\theta, \phi, \beta, v), \quad (42)$$

where $\Pi_R^\perp$ is given in Eq. (A15).

For the string part, we invite the reader to consult Appendix. The method is identical to the static case in the sense of the integration in Eq. (17), now with a modified finite-velocity permittivity. Since the resulting expressions offer no further intuition, we omit them here, but note that the constants are again set to impose the same physical boundary conditions as in the static case.

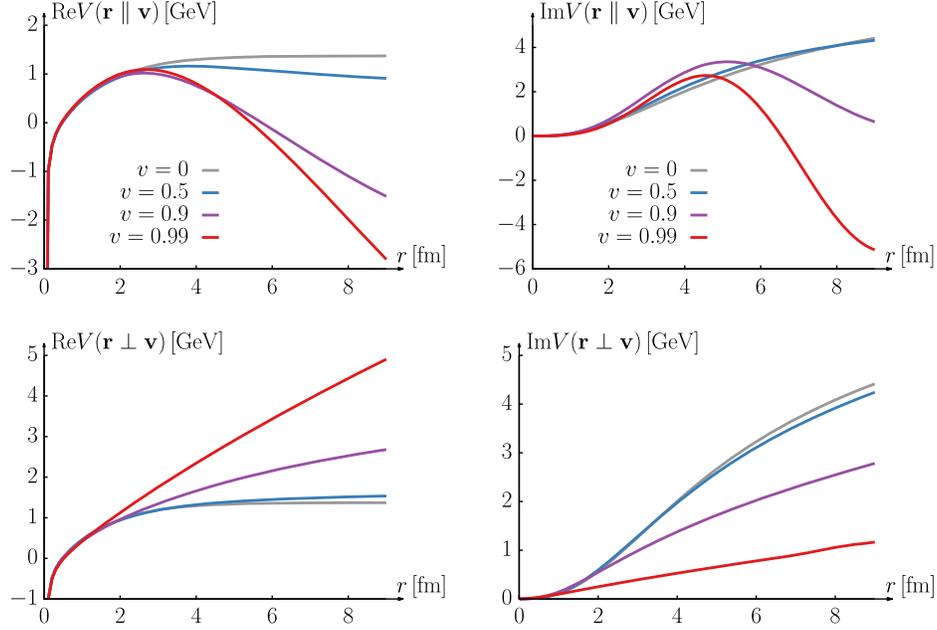

FIG. 3. The Gauss law potential model at $T = 155$ MeV for different relative velocities of the heavy quarkonium system. The top row contains the outcome for a parallel alignment of the dipole with the velocity on the bottom row for the perpendicular alignment. We show Re$V$ in the left column and Im$V$ on the right.

In Fig. 3 we show a selection of finite-velocity potentials at $T = 155$ MeV in both the parallel and the perpendicular alignments. Let us focus on the real part first. The velocity dependence for the two alignments is significantly different. While for the case $r \parallel v$ the real part is weakened with increasing $v$, the opposite happens for $r \perp v$ and one eventually recovers the Cornell potential. As had been pointed out in [44] the reason for this nontrivial behavior is the fact that the quarkonium state encounters a different effective temperature

$$T_{\text{eff}}(\theta, v) = \frac{T\sqrt{1-v^2}}{1 - v\cos(\theta)}, \quad (43)$$

depending on the magnitude and orientation of the velocity relative to the medium. The higher the velocity, the more the effective temperature deviates from its $v = 0$ value, with a hotter region in the forward direction and a cooler region in the backward direction. Since in the permittivity one integrates over the orientation $\theta$, it seems that for $r \parallel v$ the hotter region dominates, while for $r \perp v$ the colder region contributes more significantly. The fact that the real part becomes negative at large distances is not troubling, as it simply tells us that similar to the centrifugal barrier present in $P$-wave states, there is now a finite probability of the in-medium $S$-wave tunneling into an unbound configuration.

The $r \parallel v$ imaginary part of the potential behaves in a well-defined manner up to $v = 0.9c$, in that it shows nonmonotonicity but stays positive at all distances. Once Im$V$ becomes negative (in the sign convention used here) it may in principle introduce an instability when being used in a Schrödinger equation, rendering the computation unreliable. Even though the potential does show excursions of that manner for $v > 0.9c$, in practice we have not encountered any numerical difficulty when solving for the corresponding spectra presented in Sec. IV C. For $r \perp v$ the imaginary part diminishes at higher velocity, as if the quarkonium state encounters a cooler and cooler surrounding, consistent with the real part in that scenario.

In the subsequent chapters we will explore how this nontrivial modification of the potential translates into changes in the in-medium spectral functions and attempt to gain a first glimpse into production yields of heavy quarkonium in heavy-ion collisions at finite transverse momentum.

We note that in order to comprehensively capture the physics of heavy quarkonium with a finite center of momentum motion, a genuine nonequilibrium real-time approach eventually has to be developed. Efforts in this direction in the context of the open-quantum-system approach to heavy quarkonium are ongoing and will eventually incorporate information from the Gauss law model developed here.

## III. QUARKONIUM SPECTRAL PROPERTIES AT FINITE TEMPERATURE

### A. Continuum corrections

Before we can embark upon studying the in-medium properties of heavy quarkonium based on the lattice vetted

TABLE III. Bottomonium family.

|  | $\Upsilon(1S)$ | $\Upsilon(2S)$ | $\Upsilon(3S)$ | $\Upsilon(4S)$ | $\chi_{b0}(1P)$ | $\chi_{b0}(2P)$ | $\chi_{b0}(3P)$ |
|---|---|---|---|---|---|---|---|
| $m$ [GeV] | 9.4603 | 10.023 | 10.355 | 10.569 | 9.931 | 10.273 | 10.534 |
| $m^{\rm PDG}$ [GeV] | 9.4603 | 10.023 | 10.355 | 10.579 | 9.888 | 10.252 | 10.534 |
| $\langle r \rangle$ [fm] | 0.2918 | 0.5878 | 0.8697 | 1.0999 | 0.48 | 0.786 | 1.017 |
| $\bar{m}^{\rm PDG}_{B\bar{B}} - m$ [GeV] | 1.1 | 0.535 | 0.203 | −0.011 | 0.627 | 0.286 | 0.024 |

Gauss law model, we have to acknowledge the fact that the lattice data used in the previous section were not continuum extrapolated. While there is activity in the community to bring the extraction of the potential closer to the continuum limit [45], no truly extrapolated results are available today. Thus we need to manually correct for the discrepancy between the discrete lattice results and those that eventually will lead to agreement with experiment.

To this end we follow the strategy laid out in [29]. The idea is first to determine a phenomenological set of vacuum parameters that, via solving the Schrödinger equation, reproduce the masses of the known quarkonium ground state particles. In addition we will then use an appropriately rescaled version of the lattice-fitted Debye mass to implement the finite temperature effects.

The starting point is the bottomonium system, which due to the large mass of the bottom quark is the most amenable to the potential description. Furthermore, since the bottom mass is much larger than $\Lambda_{\rm QCD}$), the matching of pNRQCD with QCD can be carried out perturbatively in vacuum and one finds that the appropriate mass to use in the Schrödinger equation is the so-called renormalon subtracted mass [46]

$$m_b^{RS'} = 4.882 \pm 0.041 \text{ GeV}. \quad (44)$$

This allows us to solve a Schrödinger equation for the energy eigenstates and in turn deduce the mass of the bottomonium states, which are compared to the PDG listings [47]. With such a procedure in place, we fit the values of $\tilde{\alpha}_s$, $\sigma$, $c$ such that the masses of the lowest four $S$-wave states $\Upsilon(1S) - \Upsilon(4S)$ are reproduced to within a given accuracy. The averaged mass of the $P$-wave triplet $\chi_{b0}(1P) - \chi_{b0}(3P)$ can then be used as a cross-check. We find that

$$\tilde{\alpha}_s = 0.513 \pm 0.0024, \quad (45)$$

$$\sqrt{\sigma} = 0.412 \pm 0.0041 \text{ GeV}, \quad (46)$$

$$c = -0.161 \pm 0.0025 \text{ GeV} \quad (47)$$

are able to reproduce both the $S$-wave and the $P$-wave states very well (see Table III).

The next step is to consistently determine the only remaining unknown parameter—the charm quark mass.

Since the heavy quark potential is a universal quantity in the sense that at lowest order in pNRQCD the same expression is used for both heavy quark families, we expect the vacuum values in Eq. (45) to remain the same for charmonium. Thus by reverse engineering the procedure used above, we can "fit" the charm mass to reproduce the lowest $S$-wave states $(J/\psi, \psi')$ for our vacuum parameters. The best-fit value reads

$$m_c^{\rm fit} = 1.4692 \text{ GeV}. \quad (48)$$

Since finite mass corrections are more important for charmonium, we expect the agreement to be worse in this case, which is indeed observed in Table IV. All necessary parameters required for describing the $T=0$ physics of quarkonium within the potential approach are now set.

The next step is to consider how the lattice discretization affects the fitted values of the Debye mass parameter $m_D$. In lattice calculations the light quark masses do not take on their physical values and the chiral crossover temperature is increased. For the lattice spacings used in this study, $T_c^{\rm lat} = 172.5$ GeV. The goal is now to undo the difference between this and the physical crossover temperature by an appropriate rescaling. To this end we consider a dimensionless ratio of the lattice fitted $m_D(t)/\sqrt{\sigma(t)}$ where $t = T/T_c$. The slight dependence of the lattice string tension on the lattice spacing is taken into account in this ratio. The continuum corrected value of the Debye mass is then taken to be

$$m_D^{\rm phys}(t = T/T_c^{\rm lat}) = \frac{m_D(t)}{\sqrt{\sigma(t)}}\sqrt{\sigma^{\rm cont}}, \quad (49)$$

where $m_D(t)/\sqrt{\sigma(t)}$ is given in Table I and $\sigma^{\rm cont}$ from Eq. (45).

In order to explore the changes in the heavy quark potential at different temperatures, it is useful to also parametrize the temperature dependence of the Debye mass itself. Let us first look at the perturbative expression for the Debye mass. With dynamical quark masses $m_{u,d}$ set to zero, the leading order result for $SU(N_c)$ with $N_f$ fermions at zero baryon chemical potential is [48]

$$m_D^{\text{phys}} = Tg(\Lambda)\sqrt{\frac{N_c}{3} + \frac{N_f}{6}}, \quad (50)$$

where $\Lambda = 2\pi T$ is the renormalization scale. It is well established that the Debye mass can only be calculated up to leading order plus a logarithmic correction at next to leading order before truly nonperturbative contributions come into play [49]. For the purposes of this study, we account for this via two additional terms:

$$m_D^{\text{phys}} = Tg(\Lambda)\sqrt{\frac{N_c}{3} + \frac{N_f}{6}}$$
$$+ \frac{N_c Tg(\Lambda)^2}{4\pi}\log\left(\frac{1}{g(\Lambda)}\sqrt{\frac{N_c}{3} + \frac{N_f}{6}}\right)$$
$$+ \kappa_1 Tg(\Lambda)^2 + \kappa_2 Tg(\Lambda)^3. \quad (51)$$

TABLE IV. Charmonium family.

|  | $J/\psi$ | $\psi'$ | $\chi_{c0}(1P)$ | $\chi_{c0}(2P)$ |
|---|---|---|---|---|
| $m$ [GeV] | 3.0969 | 3.6632 | 3.5079 | 3.775 |
| $m^{\rm PDG}$ [GeV] | 3.0969 | 3.6861 | 3.4939 | 3.9228 |
| $\langle r \rangle$ [fm] | 0.565 | 1.249 | 0.672 | 1.109 |
| $\bar{m}^{\rm PDG}_{D\bar{D}} - m$ [GeV] | 0.642 | 0.076 | 0.231 | −0.036 |

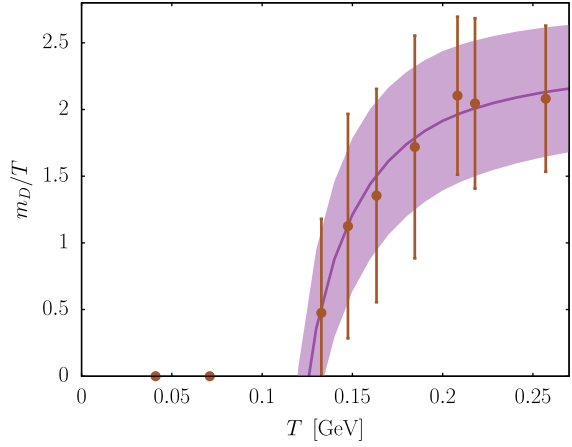

FIG. 4. Interpolated Debye mass via the HTL inspired interpolation formula (51) as a function of temperature. The error bands (purple) arise from the corresponding uncertainty in the lattice fit points.

The nonperturbative constants $\kappa_1$ and $\kappa_2$ will be fixed by performing a fit against the continuum corrected lattice results in Eq. (49). For the running coupling we utilize the four loop result given in [50] with $\Lambda_{\text{QCD}} = 0.2145$ GeV. The best fit results are

$$\kappa_1 = 0.686 \pm 0.221, \quad \kappa_2 = -0.317 \pm 0.052. \quad (52)$$

The resulting plot of Eq. (51) is shown in Fig. 4 together with the continuum corrected Debye mass points. Note that the perturbative leading order result in Eq. (50) leads to an increase in $m_D/T$ when we approach $T_c$ from above. As we eventually need to recover the $T = 0$ Cornell potential below $T_c$ the true behavior needs to exhibit a downward trend eventually, as it does in the lattice determination. This deviation from the perturbative behavior is easily captured by $\kappa_1$ and $\kappa_2$.

### B. In-medium spectral functions

Spectral functions provide the quantum field theoretical answers to questions about particle properties, i.e., their masses and decay widths. Our goal is to learn about the in-medium properties of quarkonium from an inspection of the thermal spectral functions computed from the Gauss law potential. To this end we follow the Fourier space method introduced in [51], where the following Schrödinger equation is established in describing the time evolution of the vector channel unequal-time point-split meson-meson correlator $D^>(t; \mathbf{r}, \mathbf{r}')$:

$$[\hat{H} \mp i|\text{Im}V(r)|]D^>(t; \mathbf{r}, \mathbf{r}') = i\partial_t D^>(t; \mathbf{r}, \mathbf{r}'); \quad t \gtrless 0, \quad (53)$$

with

$$\hat{H} = 2m_Q - \frac{\nabla_{\mathbf{r}}^2}{m_Q} + \frac{l(l+1)}{m_Q r^2} + \text{Re}V(r). \quad (54)$$

The correlator in frequency space is obtained from the Fourier transform

$$\tilde{D}(\omega, \mathbf{r}, \mathbf{r}') = \int_{-\infty}^{\infty} dt\, e^{i\omega t} D^>(t; \mathbf{r}, \mathbf{r}') \quad (55)$$

from which the vector channel spectral function follows by taking the limit

$$\rho^V(\omega) = \lim_{\mathbf{r},\mathbf{r}' \to 0} \frac{1}{2}\tilde{D}(\omega; \mathbf{r}, \mathbf{r}'). \quad (56)$$

A detailed discussion on how to explicitly compute the spectral functions from Eq. (53), both formally and practically, can be found in Appendix A of [51]. Furthermore, the pseudoscalar and axial vector channels were shown not to include any qualitatively new structures, with

$$\rho^P \simeq -\frac{1}{3}\rho^V, \quad \rho^{A^0} \simeq -\frac{1}{3}\rho^V, \quad \rho^{\mathbf{A}} \simeq 2\rho^V. \quad (57)$$

In this study we focus on the $S$-wave quarkonium spectral functions and show in Fig. 5 the results from the improved Gauss law model for both the bottomonium (top) and the charmonium (bottom) vector channel. The gray dashed lines correspond to the vacuum spectral functions, which exhibit three well-defined bound states for bottomonium and two for charmonium below threshold. Qualitatively similar to what has been reported in the literature [29], we find characteristic changes as temperature increases. Both a broadening of the peaks and a shift of their central value to smaller frequencies is observed. We note that the strength of the change is clearly ordered with the vacuum binding energy of each individual state, where $E_{\rm bind}$ is defined from the distance between the spectral peak and the threshold. Just as intuition predicts, the more deeply the state is bound the less susceptible it is to medium effects. This behavior is observed consistently in potential based computations and is also in agreement with recent studies of directly reconstructed bottomonium and charmonium spectral functions from lattice QCD [15].

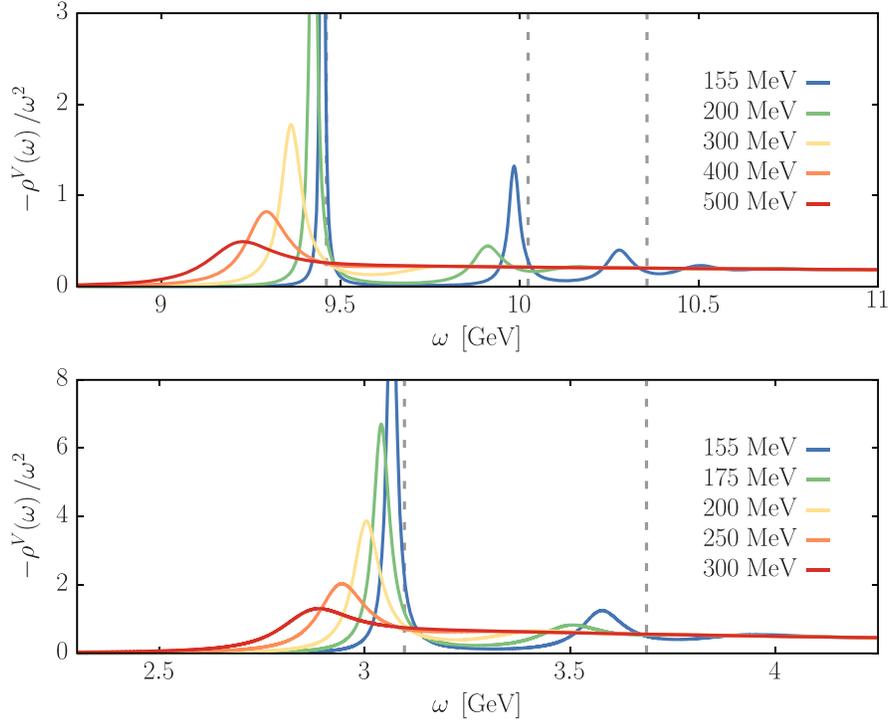

FIG. 5. Representative examples of in-medium S-wave spectral functions for vector channel bottomonium (top) and charmonium (bottom). The gray dashed lines denote the $T = 0$ stable bound states below the threshold, calculated from the ($T = 0$) Schrödinger equation after calibrating the string breaking radius to 1.25 fm. At finite temperature a characteristic broadening and shifting to lower frequencies is observed, consistent with previous potential based studies.

To more quantitatively explore the in-medium properties we can consult scattering theory. If a narrow resonance pole lies close to the real frequency axis, then its spectrum can be described by a Breit-Wigner distribution. On the other hand, if it features a significant decay width we may employ a skewed Breit-Wigner of the form

$$\rho(\omega \approx E) = C\frac{(\Gamma/2)^2}{(\Gamma/2)^2 + (\omega-E)^2} + 2\delta\frac{(\omega-E)\Gamma/2}{(\Gamma/2)^2 + (\omega-E)^2}$$
$$+ C_1 + C_2(\omega - E) + \mathcal{O}(\delta^2), \qquad (58)$$

which is able to disentangle the bound state signal from the background continuum. Here, $E$ denotes the energy of the resonance, $\Gamma$ its width, and $\delta$ the phase shift. The constant terms $C_1$ and $C_2$ model artifacts beyond the spectral peak we are interested in.

For completeness we plot in Figs. 6 and 7 the in-medium masses, as well as the threshold behavior for charmonium and bottomonium, respectively. The values obtained here, while qualitatively consistent with studies based on the legacy formulation of the Gauss law model [29], show a somewhat higher stability of in-medium quarkonium. The reason lies in the extrapolation ambiguity of the fitted Re$V$ to distances $r > 1$ fm, where the lattice data are currently unable to constrain the functional form. High precision lattice determinations of Re$V$ would be required to resolve this ambiguity.

## IV. APPLICATIONS TO HEAVY-ION COLLISIONS

The computation of the in-medium spectral functions has already provided us with vital insight on the properties of heavy quarkonium in thermal equilibrium. The question, however, remains of how to connect this information to actual measurements carried out in heavy-ion collision experiments. Contrary to light mesons, where a direct link exists between in-medium spectral functions and the measured decay leptons, for heavy quarkonium we do not measure their thermal decay. Instead, at hadronization a number of vacuum states are created whose decay is recorded long after the QGP has ceased to exist. Thus we need to translate all in-medium information into abundances of vacuum states to connect to experiment.

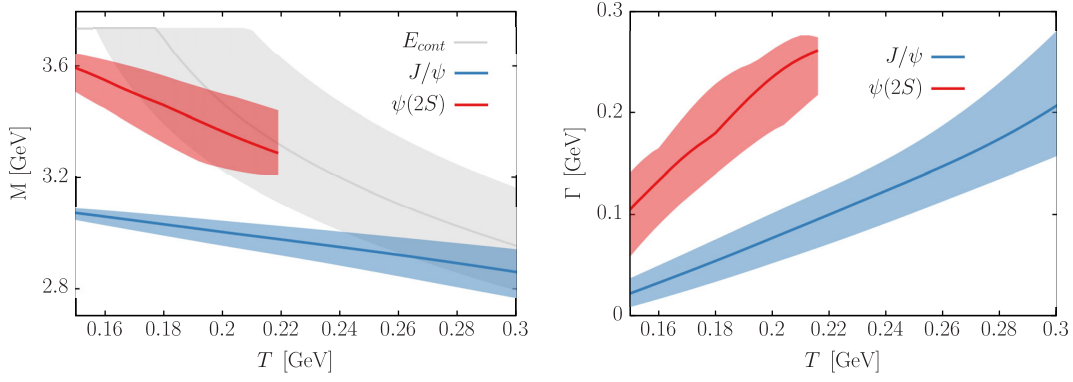

FIG. 6. Thermal mass (left) and spectral width (right) of charmonium as a function of temperature. The error bands denote the Debye mass uncertainty arising from the fitting procedure. The continuum threshold energy on the left figure is defined as $\mathrm{Re}V(r \to \infty)$.

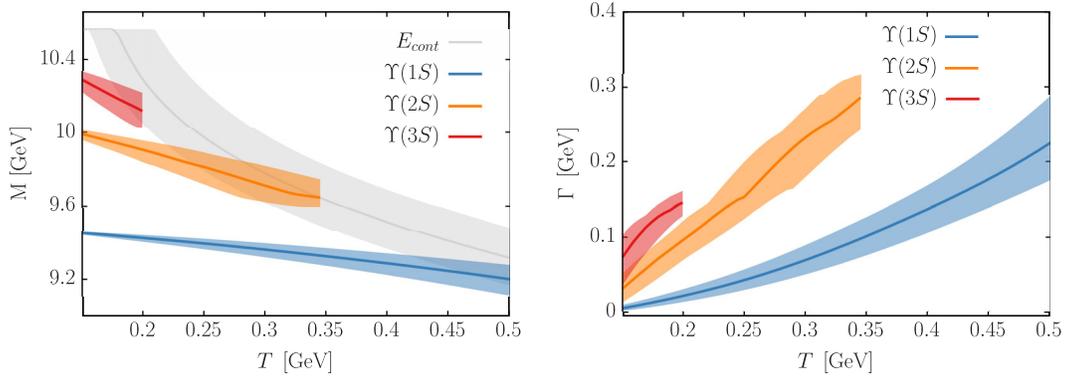

FIG. 7. Thermal mass (left) and spectral width (right) of bottomonium as a function of temperature. The error bands denote the Debye mass uncertainty arising from the fitting procedure. The continuum threshold energy on the left figure is defined as $\mathrm{Re}V(r \to \infty)$.

In previous studies the Gauss law model has been deployed to investigate the production of quarkonium in heavy-ion collisions in the simplest possible scenario, where the quarks are almost at rest with the surrounding medium. In addition, since lattice QCD simulations have not yet produced results for the heavy-quark potential at finite baryon density, only predictions for highest energy LHC collisions have been presented.

Here we will go beyond these results by modeling heavy quarkonium production both at finite baryochemical potential, relevant for lower-energy heavy-ion collisions at RHIC, FAIR, and NICA, and at the finite transverse momentum.

### A. $\psi'$ to $J/\psi$ production ratio

Among the currently most highly sought observables at the LHC heavy-ion program is the ratio of $\psi'$ to $J/\psi$ produced in Pb-Pb collisions. In contrast to the nuclear modification factor $R_{AA}$ for each individual species, this ratio promises to be highly discriminatory between different phenomenological models for quarkonium production in heavy-ion collisions. Our goal is therefore to estimate this ratio.

Following the ideas laid out in [29], we utilize the in-medium spectral functions computed in the preceding sections and will assume kinetic thermalization of the charm quarks in the late stages of a collision. The idea is to convert in a meaningful fashion the in-medium spectral information about $\psi'$ and $J/\psi$ into the number of produced vacuum states. The assumption behind this step is that of an instantaneous freeze-out, where at $T = T_c$ the in-medium particles abruptly change into vacuum particles.

To translate in-medium spectral peaks into vacuum states we consider the relation between the dilepton emission rate and the in-medium spectral function [52]:

$$\frac{\mathrm{d}R_{\ell\bar{\ell}}}{\mathrm{d}P^4} = -\frac{Q_q \alpha_e^2}{3\pi^2 P^2} n_B(p_0) \rho^V(P). \quad (59)$$

Here, $n_B$ denotes the Bose-Einstein factor, $Q_q$ the electric charge of the heavy quark in units of $e$, and $\alpha_e$ the QED coupling constant. The four-momentum is denoted $P = (p_0, \mathbf{p})$, and the finite mass of the leptons has been neglected. Equation (59) relates the weighted area under the in-medium peak to the rate of dileptons emitted from that state. Thus, integrating the right-hand side above gives the number of in-medium lepton pairs produced by each of the states. The prefactors drop out if only the ratio is computed, leaving

$$R_{\ell\bar{\ell}} \propto \int \mathrm{d}p_0 \mathrm{d}^3\mathbf{p} \frac{\rho^V(P)}{P^2} n_B(p_0). \quad (60)$$

Let us emphasize again that this quantity is not what is measured in experiment. Since the plasma is diluted away long before the charmonium states decay, the number of dileptons eventually measured originate from the vacuum-state remnants of the in-medium structures observed here.

Thus we must project the states corresponding to the finite temperature peaks onto the ($T = 0$) vacuum states. The computation proceeds as follows. At leading order the vector channel spectral function $\rho^V$ depends only on $P^2 = p_0^2 - \mathbf{p}^2$; after performing a change of variables to $\omega = \sqrt{p_0^2 - \mathbf{p}^2}$, Eq. (60) becomes

$$R_{\ell\bar{\ell}} \propto \int \mathrm{d}\omega \mathrm{d}^3\mathbf{p} \frac{\rho^V(\omega)}{\omega^2} n_B\left(\sqrt{\omega^2 + \mathbf{p}^2}\right) \frac{\omega}{\sqrt{\omega^2 + \mathbf{p}^2}}. \quad (61)$$

In this expression, the contribution from each bound state arises from the corresponding peak area in $\rho^V(\omega)/\omega^2$. We fit each peak structure with the skewed Breit-Wigner in Eq. (58), thus allowing the different contributions to be distinguished and the thermal mass $M_n$ and width of each to be ascertained. Now, the projection onto the vacuum states is implemented by writing $\rho^V(\omega)/\omega^2 = A_n \, \delta(\omega - M_n)$; that is, we allow the in-medium states to collapse into delta peaks that represent the vacuum states, while retaining the peak area $A_n$ to account for the different contributions. We have also confirmed numerically that it is indeed possible to approximate the Breit-Wigner peak by a delta function in this manner. Imposing this on Eq. (61) and carrying out the now trivial $\omega$ integral then gives

$$R_{\ell\bar{\ell}}^{\psi_n} \propto A_n \int \mathrm{d}^3\mathbf{p}\, n_B\left(\sqrt{M_n^2 + \mathbf{p}^2}\right) \frac{M_n}{\sqrt{M_n + \mathbf{p}^2}}. \quad (62)$$

Switching to spherical coordinates in momentum then leaves only a $\mathrm{d}p$ integral that can be easily evaluated numerically.

Finally, in order to obtain the total number density we must divide by the electromagnetic decay rate of the vacuum state, which is proportional to the square of the wave function at ($r = 0$) divided by the square of the mass of the state [53]. These values we calculate ourselves using the corresponding wave functions from the spectroscopic fitting performed in Sec. III A. The final expression is

$$\frac{N_{\psi'}}{N_{J/\psi}} = \frac{R_{\ell\bar{\ell}}^{\psi'}}{R_{\ell\bar{\ell}}^{J/\psi}} \cdot \frac{M_{\psi'}^2 |\psi_{J/\psi}(0)|^2}{M_{J/\psi}^2 |\psi_{\psi'}(0)|^2}. \quad (63)$$

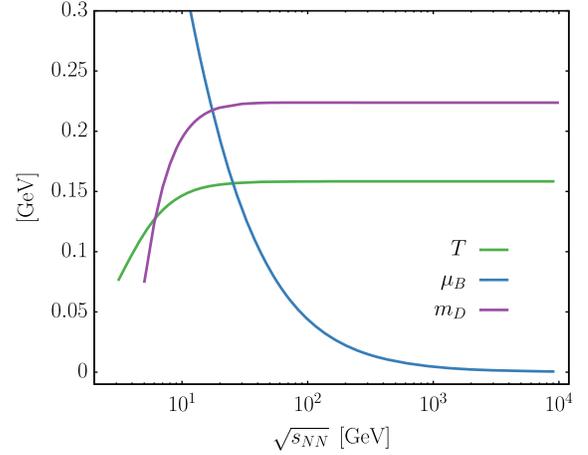

FIG. 8. The most recent values of temperature (green curve) and baryochemical potential (blue curve) extracted from the statistical model of hadronization for different beam energies $\sqrt{s_{NN}}$. To estimate the $\psi'/J/\psi$ ratio we translate these values into a corresponding Debye mass parameter $m_D(T, \mu_B)$ for the Gauss law model (purple curve).

### B. Finite $\mu_B$ phenomenology

In order to make use of Eq. (63) in connecting to experiment, we require a prescription to evaluate our Gauss law potential model (and hence the resulting spectral functions) at a given center-of-mass energy.

The strategy here is twofold. First, we note that the successful statistical hadronization model provides a well-established scheme with which to estimate the thermal parameters of the produced medium at chemical freeze-out resulting from a collision at a given $\sqrt{s_{NN}}$. Starting from the grand canonical partition function of known hadrons, one is able to reproduce the measured yields by adjusting the three parameters of the model: collision volume, temperature, and baryochemical potential ($\mu_B$). The most recent values [54] are

$$T(\sqrt{s_{NN}}) = \frac{158 \text{ MeV}}{1 + \exp(2.60 - \ln(\sqrt{s_{NN}})/0.45)}, \quad (64)$$

$$\mu_B(\sqrt{s_{NN}}) = \frac{1307.5 \text{ MeV}}{1 + 0.288\sqrt{s_{NN}}}, \quad (65)$$

where $\sqrt{s_{NN}}$ is the dimensionless numerical value of the center-of-mass energy measured in GeV. These are plotted in Fig. 8. In this model the freeze-out temperature quickly asymptotes to the limiting temperature of 158 MeV while $\mu_B$ drops monotonously to almost zero at high collision energies such as those probed at RHIC and LHC. Second, note that the medium effects in our potential model are captured entirely by the value of the Debye mass $m_D$. Equation (51) gives our interpolated continuum-corrected expression at $\mu_B = 0$, denoted now as $m_D(T, \mu_B = 0)$. We now postulate how to extend that formula into the realm of the finite baryochemical potential. At leading order, the Debye mass can be directly calculated perturbatively at finite baryochemical potential [48]. We propose to add this $\mu_B$ dependence to the temperature-only dependence of Eq. (51). This is valid only for small values of $\mu_B$ and at leading order. On the other hand, at very large values of $\mu_B$ the chemical potential itself becomes the only relevant scale, and we expect from dimensional grounds that it again enters $m_D$ linearly. Thus our modeling assumption is to adopt the following at all values of $\mu_B$:

$$m_D(T, \mu_B) = \sqrt{m_D^2(T, 0) + T^2 g^2 \frac{N_f}{18\pi^2} \frac{\mu_B^2}{T^2}}. \quad (66)$$

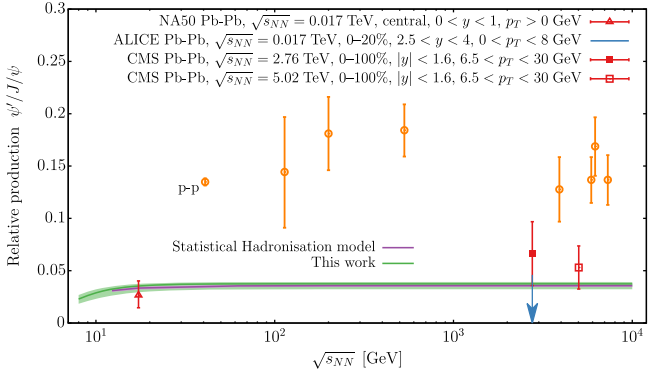

FIG. 9. The prediction of this work (green curve) for the relative production yield of $\psi'$ to $J/\psi$. We also include the statistical hadronization model prediction [54] (purple curve) and experimental data measured by the NA50 [55], ALICE [56], and CMS [57,58] Collaborations (red symbols) for Pb-Pb collisions as well as the $pp$ baseline [54,59] (orange symbols).

Here, the renormalization scale is also modified to $\Lambda = 2\pi\sqrt{T^2 + \mu_B^2/\pi^2}$. It is important to point out that progress has been made in computing the Debye mass perturbatively at next-to-leading order in the chemical potential [60], which would modify Eq. (66). However, as a naive first approximation we employ Eq. (66) in its current form to extend the Gauss law parametrization into the finite baryochemical potential regime. Additionally, in the absence of reliable lattice data at the nonzero quark chemical potential, we hold the nonperturbative constants $\kappa_1$ and $\kappa_2$ in $m_D(T, 0)$ the same as in Eq. (52). Under these assumptions we may explore the domain of finite $\mu_B$ and via the information provided from the statistical model, also at different $\sqrt{s_{NN}}$ [i.e., via Eq. (65)].

With all ingredients in place, we may now proceed with calculating the $\psi'$ to $J/\psi$ ratio over a range of center-of-mass energies. The results from this entire procedure are plotted in Fig. 9, as well as a comparison with the statistical hadronization model prediction and the most up-to-date experimental data for Pb-Pb collisions and the $pp$ baseline. This extends previously available computations, valid only at the highest beam energies, to those relevant for NICA and FAIR.

Our analysis is based on in-medium spectral functions and is independent from that performed by the statistical hadronization model. The only information shared among the two are the values for $T$ and $\mu_B$ extracted from the yields of light hadrons. We find as expected from the previous Gauss law studies that the results at vanishing $\mu_B$ lie very close to the prediction from the statistical model. In the lattice fits carried out using the new and improved Gauss law model we have been more conservative in the estimation of our uncertainties, which is why the present results are fully compatible with the statistical model. Note that while different in form, both approaches share that they consider a fully kinetically thermalized scenario. A good agreement between the two and the experimental data thus further supports the interpretation that charmonium at LHC has reached a significant degree of kinetic equilibration with its surrounding.

We find that extending our Gauss law model to finite $\mu_B$, i.e., lower $\sqrt{s_{NN}}$, the agreement with the statistical model persists. Even though our assumptions to do so were rather crude, they lead to both a qualitatively and an even quantitatively very similar trend for $\psi'/J/\psi$. On the other hand, the full validity of these results when compared to experimentally measured data is somewhat questionable. Whether charmonium can be considered as kinetically thermalized in collisions as low as $\sqrt{s_{NN}} \sim 40$ MeV, for example as carried out at RHIC, remains to be seen, in particular in light of the difficulties of measuring a finite elliptic flow for $J/\psi$ there.

### C. Finite transverse momentum

Not long after the initial interest in quarkonium as a probe of the QGP, the first works began to appear that considered how the naive Coulombic Debye screening description could be extended to account for quarkonium moving with a finite velocity [61]. Recent years have seen a revival of interest in this direction, with new approaches employing modern effective field theory techniques to tackle the problem [43,44,62,63].

The phenomenological motivations are clear; quarkonia produced in heavy-ion collisions traverse the hot medium before being measured with finite transverse momenta $p_T$, and accounting for this may lead to qualitatively new QGP phenomena such as the formation of wakes [64–66]. In order to take a first step toward realistic phenomenology based on the finite velocity Gauss law model constructed in Sec. II F, we require a prescription for converting from a general medium velocity $\mathbf{v}$ to the transverse momenta $p_T$ commonly measured in heavy-ion collisions. Such a prescription has been described in [63], which we briefly review here. Consider a heavy quarkonium traversing the QGP with a momentum $P^\mu = (p^0, \mathbf{p})$ measured in the lab frame. The QGP will also have a velocity in that frame, denoted **w**, and it is the relative motion that needs to be estimated and subsequently employed as **v** in the Gauss law potential. Assuming a central collision and constant **w** throughout, a typical value for the LHC is $w \sim w_\perp \sim 0.66$. Further assuming a thermalized and isotropic system, the modulus of the relative velocity will be given as

$$v = \sqrt{1 - \frac{(1-w^2)M^2}{M^2 - 2p^0 \mathbf{w}\cdot\mathbf{p} + (\mathbf{w}\cdot\mathbf{p})^2 + \mathbf{p}^2}}, \quad (67)$$

where $M$ is the heavy quarkonium mass. Note that this depends on the angle $\varphi$ between the QGP velocity **w** and the quarkonium three-momentum **p**.

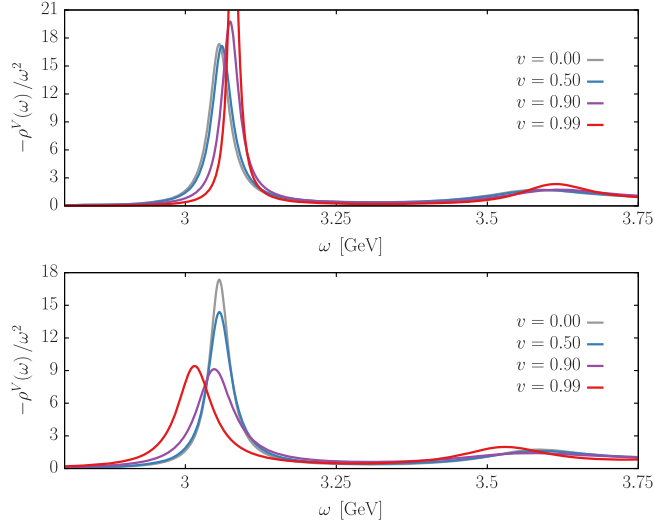

FIG. 10. Representative examples of charmonium finite velocity spectra, for both the parallel (top) and the perpendicular (bottom) cases, at $T = 158$ MeV and $\mu_B = 0$. The gray curve shows the static result. Note that the y-axis scale varies slightly between the two panels.

Following the same procedure as the preceding sections, we may now calculate finite velocity spectral functions.

Some representative examples are shown for charmonium in Fig. 10. The qualitative difference in the potential for parallel (top) and perpendicular (bottom) alignments manifests itself also in the behavior of the spectral function. In the perpendicular case, going to higher velocities is reminiscent of increasing temperatures in that the peak, generally speaking, is broadened and shifted to lower frequencies. Note, however, that this trend is eventually reversed at ultrarelativistic velocities. When investigating the physics of bottomonium at finite velocity in an EFT picture [44], a similar phenomenon was encountered with increasing velocity leading to spectral modifications similar to an increased temperature. In contrast, the parallel case exhibits a shift to higher frequencies with little or no effect on the width.

We may also use the procedure of the preceding sections to estimate the $\psi'$ to $J/\psi$ production ratio at finite velocity and through Eq. (67) can relate this to a simplified description of a heavy-ion collision. In Eq. (67) we replace **p** with $\mathbf{p}_T$ and further assume a uniform distribution over $\varphi$ (see left panel of Fig. 11) before taking the mean. The results of this method at large center-of-mass energies are shown in the right panel of Fig. 11. We find that the effect of the finite center of momentum motion on the quarkonium state in the fully thermal Gauss law model is moderate. Up to the $p_T = 25$ GeV considered here, we find an 11% increase in the production ratio for a parallel alignment of the dipole and up to 17% for a perpendicular alignment. We have to keep in mind that the values obtained here rely on many simplifying assumptions, for example that the expansion of the fireball was taken to be isotropic with a constant velocity. What may help us is that if the production of quarkonium really is dominated by the physics around the freeze-out, as suggested by the statistical model of hadronization, then we indeed only need to track the shells of the QCD medium, which at a given moment are close to $T = T_c$. How well the model presented here captures the physics of quarkonium in a heavy-ion collision will be testable at the upcoming Run 3 at the LHC, where the first accurate experimental values for $\psi'/J/\psi$ are expected to become available.

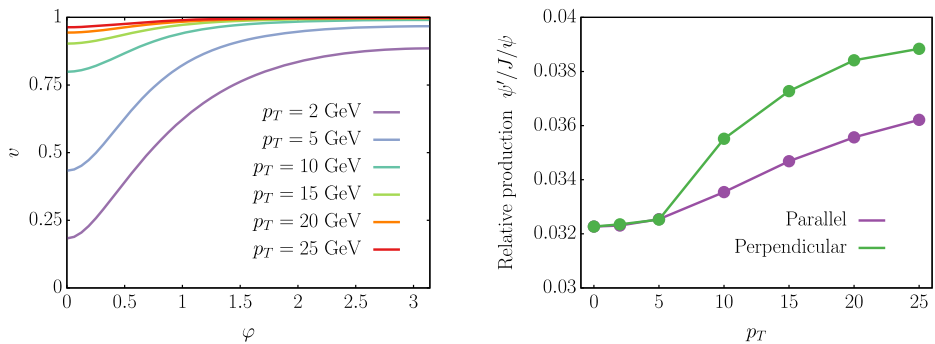

FIG. 11. (Left) The angular dependence of the relative velocity for different transverse momenta, given by Eq. (67). The distribution is symmetric around $\varphi = \pi$. (Right) The prediction of this work for the relative production yield of $\psi'$ to $J/\psi$ particles as a function of transverse momentum $p_T$ for both the parallel (purple curve) and the perpendicular (green curve) alignments, at ($T = 158$ MeV and $\mu_B = 0$ corresponding to a large center-of-mass energy.

A recent estimate [67] of the $p_T$ dependence of the ratio, combining the statistical model with a more realistic evolution of the temperature profile in the collision center and careful separation of the core and corona of the collision, predicts a strong change up to a factor of 3–4.

## V. CONCLUSION

Heavy quarkonium is a vital tool in developing our understanding of strongly interacting matter and connecting experimental results to the fundamental theory of QCD. On the one hand, progress in effective field theories has demonstrated rigorously the validity of describing their in-medium behavior with an effective potential in an appropriately defined Schrödinger equation; on the other, advancements in lattice techniques continue to provide nonperturbative results against which such approximations can be checked.

The main conceptual result of this work is a rigorously derived model for the in-medium heavy quark potential based on the generalized Gauss law in linear response theory. It describes the in-medium modification of the vacuum Cornell potential by self-consistently incorporating a weakly coupled medium described by the HTL permittivity. The resulting analytic expressions depend only on a single temperature dependent parameter and are able to reproduce the lattice results for the real part of the potential even in the nonperturbative regime close to $T_c$. The string imaginary part showed an unphysical logarithmic divergence, which we attribute to the equally unphysical unending linear rise of the vacuum Cornell potential. By considering the presence of string breaking it is possible to regularize this artifact, and we were able to give physically sound predictions for the imaginary part of the potential that qualitatively matched the lattice data. The presented work has improved on the conceptual clarity and technical robustness of the Gauss law model compared to other potential models described in the literature. In preparation of upcoming high resolution lattice QCD data for the interquark potential, the straightforward extension of the Gauss law model with a running coupling has been discussed.

Using a combination of weak-coupling computations and dimensional analysis, we introduced an extension of the Gauss law model to finite baryochemical potential. The extension to quarkonium moving relative to the QCD medium required us to consider two separate alignments of the quark-antiquark dipole with respect to the velocity when computing the corresponding in-medium permittivity. The resulting expressions for the in-medium potential, while lengthy, could be provided in explicit form. Both extensions of the model are required to step toward describing phenomenologically relevant scenarios in heavy-ion collisions, which are not yet amenable to direct lattice QCD simulations.

A continuum correction on the vacuum parameters, as well as $m_D$, was performed that allowed physically realistic spectral functions to be computed in the S-wave channel, which formed the basis of our phenomenological investigation. Similar to previous studies based on the previous Gauss law model, we find the characteristic broadening and shifting of spectral features to lower frequencies with increasing temperature. The strength of the in-medium modification is hierarchically ordered with the vacuum binding energy. A skewed Breit-Wigner was fitted to each resonance peak in order to obtain quantitative results for in-medium masses and thermal width, which are consistent with previous studies.

The first phenomenological result of this work lies in extending the calculation of the $\psi'$ to $J/\psi$ production yield to finite baryochemical potential and subsequently to lower beam energies relevant for NICA and FAIR. By assuming an instant freeze-out at around the chiral crossover temperature, we found an excellent agreement with the statistical hadronization model, and our prediction aligned with the latest results from ALICE and CMS to within the experimental errors. As our approach, based on in-medium spectral functions, is largely independent of the statistical model of hadronization but shares the idea of a fully kinetically thermalized quarkonium, the agreement corroborates the interpretation of charm quarks becoming equilibrated in the hot fireball before transitioning into vacuum states at the freeze-out boundary.

The second phenomenological result is our estimate of the change in the $\psi'$ to $J/\psi$ production yield for finite transverse momentum. We find increases between 11% and 17% for an increase in $p_T$ from zero to 25 GeV, which is moderate compared to predictions based on the statistical model which foresees an increase by a factor of 3–4.

We are confident that the availability of this cleanly derived and lattice-vetted complex potential model will be of use to the quarkonium phenomenology community. The Gauss law model described here is future proof, as it is ready to accommodate the upcoming high precision and high resolution lattice data on the interquark potential, where for example a running coupling will be relevant. While in this study we were only able to access the information contained in thermal in-medium spectral functions, we are looking forward to seeing the complex potential inform simulations in the open-quantum-systems framework for heavy quarkonium, where a more detailed analysis of the in-medium real-time evolution and recombination dynamics at freeze-out are possible.

## ACKNOWLEDGMENTS

We are grateful to Anton Andronic for providing the statistical hadronization model predictions. This work is part of and supported by the DFG Collaborative Research Centre "SFB 1225 (ISOQUANT)."

## APPENDIX: HEAVY QUARK POTENTIAL AT FINITE VELOCITY

The work of [63] first investigated nonrelativistic QED bound states moving with finite velocity in a background thermal bath via a rigorous effective field theory approach. This was then generalized to QCD and heavy quarkonium in [44], which forms the starting point of our analysis. The general framework is as follows. We assume that the QCD plasma is in thermal equilibrium at temperature $T$ and consider a reference frame in which the medium moves with velocity **v** with respect to the heavy quark bound state $(Q\bar{Q})$ at rest. This frame has been used successfully in the past [68]. The particle distribution functions are given by

$$f(\beta^\mu P_\mu) = \frac{1}{\exp[\beta^\mu P_\mu] \pm 1}, \tag{A1}$$

where the plus (minus) sign refers to bosons (fermions) and

$$\beta^\mu = \frac{\gamma}{T}(1, \mathbf{v}) = \frac{u^\mu}{T}. \tag{A2}$$

Here, $\gamma = 1/\sqrt{1-v^2}$ is the Lorentz factor with $v = |\mathbf{v}|$ and the four-momentum is $P = (p_0, \mathbf{p})$. The study of a bound state in a moving thermal medium is equivalent to studying a bound state in nonequilibrium field theory [69]; in such a formalism the Bose-Einstein or Fermi-Dirac distributions are generalized, which in our case are given by the boosted versions in Eq. (A1). For a thermal medium of massless (anti)particles, nonequilibrium field theory gives

$$\beta^\mu P_\mu = p\frac{1 - v\cos(\theta)}{T\sqrt{1-v^2}}, \tag{A3}$$

where $p = |\mathbf{p}|$ and $\theta$ denotes the angle between **p** and **v**. The distribution functions in Eq. (A1) then become

$$f(p, T, \theta, v) = \frac{1}{\exp[p/T_{\text{eff}}(\theta, v)] \pm 1}, \tag{A4}$$

where the effective temperature is defined

$$T_{\text{eff}}(\theta, v) = \frac{T\sqrt{1-v^2}}{1 - v\cos(\theta)}. \tag{A5}$$

Intuitively, Eq. (A5) can be understood as a Doppler effect. For $v \ll 1$ it is shown in [44] that $T_{\text{eff}} \sim T$ for all directions $\theta$, while for $v \sim 1$ the temperature felt by the bound state varies more significantly. The new scales introduced by considering a thermal medium can be understood via light-cone coordinates by defining a maximum $(T_+)$ and minimum $(T_-)$ measurable temperature with $T_- < T < T_+$.

The subsequent discussion assumes $T_- \sim T_+$, which is strictly not true as $v \to 1$; however, in [63] it was found that correct results were obtained for QED by a simple extrapolation.

The authors in [44] then proceed with an inspection of the hierarchy of scales in the formalism outlined above. The details will not be included here; however, the argumentation follows the construction of pNRQCD$_{\text{HTL}}$. In the regime $T \gg 1/r \gg m_D \gg E_{\text{binding}}$, one may employ the HTL real-time formalism and extend the computation of the heavy quark potential at finite temperature to include the presence of a moving thermal bath. This was first performed for the real Coulombic part in [61,65] before being extended to the newly understood imaginary part in [63,70]. More recently, these results were combined with the linear response ansatz in order to model the in-medium and finite-velocity modifications to the string as well as the Coulombic part of the Cornell potential [43].

We will now briefly review the real-time HTL calculation, following the procedure in [43], before applying our model prescription. The physical component of the longitudinal component of the gluon propagator can be written in terms of the corresponding retarded, advanced, and symmetric propagators,

$$D_{11}(p_0 = 0, \mathbf{p}, u) = \frac{1}{2}[D_R(\mathbf{p}, u) + D_A(\mathbf{p}, u) + D_S(\mathbf{p}, u)], \tag{A6}$$

where each propagator can be obtained from its corresponding self-energy. Note that we have made explicit the velocity dependence, since these represent different quantities to those discussed previously. In this framework, the symmetric propagator is given as

$$D_S(\mathbf{p}, u) = \frac{\Pi_S(\mathbf{p}, u)}{2i\text{Im}\Pi_R(\mathbf{p}, u)}[D_R(\mathbf{p}, u) - D_A(\mathbf{p}, u)], \tag{A7}$$

where $\Pi_{R(S)}(\mathbf{p}, u)$ is the retarded (symmetric) self-energy. In the frame where the bound state is at rest, the retarded self-energy can be parametrized as

$$\Pi_R(\mathbf{p}, u) = a(z) + \frac{b(z)}{1-v^2} \tag{A8}$$

with

$$a(z) = \frac{m_D^2}{2}\left[z^2 - (z^2 - 1)\frac{z}{2}\log\left(\frac{z+1+i\epsilon}{z-1+i\epsilon}\right)\right], \tag{A9}$$

$$b(z) = (z^2 - 1)\left[a(z) + m_D^2(z^2 - 1)\right.$$
$$\left.\times\left(1 - \frac{z}{2}\log\left(\frac{z+1+i\epsilon}{z-1+i\epsilon}\right)\right)\right] \tag{A10}$$

for

$$z = \frac{P \cdot u}{\sqrt{(P \cdot u)^2 - P^2}}\bigg|_{p_0=0}. \quad (A11)$$

The $Q\bar{Q}$ dipole is always considered to lie along the $z$ axis, which gives rise to two separate alignments. First, if the velocity direction is parallel to the axis of the dipole, Eq. (A11) becomes

$$z_\parallel = \frac{v\cos(\theta)}{\sqrt{1 - v^2\sin^2(\theta)}}, \quad (A12)$$

where $\theta$ is now simply the polar angle of the momentum vector. One should keep in mind that the momenta here are associated with the mediating gluons and thus align with the dipole direction. In the second case the velocity of the medium lies in the $x - y$ plane and makes an angle $\beta$ with the $x$ axis. Labeling the azimuthal angle of the momentum vector as $\phi$, Eq. (A11) then becomes

$$z_\perp = \frac{v\sin(\theta)\cos(\phi - \beta)}{\sqrt{1 - v^2 - v^2\sin(\theta)\cos(\phi - \beta)}}. \quad (A13)$$

After some manipulations, Eqs. (A8)–(A13) can be combined to give the complex retarded gluon self-energy in the parallel and perpendicular directions:

$$\Pi_R^\parallel(\theta, v) = \frac{m_D^2}{2}\left[\frac{2 - 2v^2 - v^4\cos^2(\theta)\sin^2(\theta)}{(1 - v^2\sin^2(\theta))^2} - \frac{(2 + v^2\sin^2(\theta))(1 - v^2)v\cos(\theta)}{2(1 - v^2\sin^2(\theta))^{5/2}}\log\left(\frac{v\cos(\theta) + \sqrt{1 - v^2\sin^2(\theta)}}{v\cos(\theta) + \sqrt{1 + v^2\sin^2(\theta)}}\right)\right], \quad (A14)$$

$$\Pi_R^\perp(\theta, \phi, \beta, v) = \frac{m_D^2}{2}\left[\frac{2 - 2v^2 - v^4\sin^2(\theta)\cos^2(\phi - \beta)(1 - \sin^2(\theta)\cos^2(\phi - \beta))}{(1 - v^2 + v^2\sin^2(\theta)\cos^2(\phi - \beta))^2}\right.$$
$$- \frac{2 + v^2 - v^2\sin^2(\theta)\cos^2(\phi - \beta)(1 - v^2)v\sin(\theta)\cos^2(\phi - \beta)}{(1 - v^2 + v^2\sin^2(\theta)\cos^2(\phi - \beta))^{5/2}}$$
$$\left.\times \log\left(\frac{v\cos(\phi - \beta)\sin(\theta) + \sqrt{1 - v^2 + v^2\sin^2(\theta)\cos^2(\phi - \beta)}}{v\cos(\phi - \beta)\sin(\theta) - \sqrt{1 - v^2 + v^2\sin^2(\theta)\cos^2(\phi - \beta)}}\right)\right]. \quad (A15)$$

With these expressions, the retarded propagator is obtained via

$$D_R^{\parallel(\perp)}(\mathbf{p}, u) = \frac{1}{p^2 + \Pi_R^{\parallel(\perp)}(\mathbf{p}, u)}. \quad (A16)$$

Furthermore, the advanced self-energy can be obtained with the relation

$$D_A^{\parallel(\perp)}(\mathbf{p}, u) = (D_R^{\parallel(\perp)}(\mathbf{p}, u))^*. \quad (A17)$$

Similarly, one can calculate the symmetric self-energy for both cases [63]. The result is

$$\Pi_S^\parallel(\mathbf{p}, u) = i\frac{2\pi m_D^2 T(1 - v^2)^{3/2}(1 + \frac{v^2}{2}\sin^2(\theta))}{p(1 - v^2\sin^2(\theta))^{5/2}} \quad (A18)$$

and

$$\Pi_S^\perp(\mathbf{p}, u) = i\frac{2\pi m_D^2 T(1 - v^2)^{3/2}(1 + \frac{v^2}{2} + \frac{v^2}{2}\sin^2(\theta)\cos^2(\phi - \beta))}{p(1 - v^2 + v^2\sin^2(\theta)\cos^2(\phi - \beta))^{5/2}}. \quad (A19)$$

We now have all of the ingredients to assemble Eq. (A7). From Eq. (A17) we attain

$$D_R^{\parallel(\perp)}(\mathbf{p}, u) - D_A^{\parallel(\perp)}(\mathbf{p}, u) = \frac{1}{p^2 + \Pi_R^{\parallel(\perp)}(\mathbf{p}, u)} - \frac{1}{p^2 + (\Pi_R^{\parallel(\perp)}(\mathbf{p}, u))^*} = \frac{2i\text{Im}\Pi_R^{\parallel(\perp)}(\mathbf{p}, u)}{[p^2 + \Pi_R^{\parallel(\perp)}(\mathbf{p}, u)][p^2 + (\Pi_R^{\parallel(\perp)}(\mathbf{p}, u))^*]}. \quad (A20)$$

Thus, the symmetric propagator for the parallel case is found as

$$D_S^{\parallel}(\mathbf{p}, u) = \frac{-2\pi i m_D^2 T (1-v^2)^{3/2}(2 + v^2 \sin^2(\theta))}{2p(1 - v^2 \sin^2(\theta))^{5/2}[p^2 + \Pi_R^{\parallel}(\mathbf{p}, u)][p^2 + (\Pi_R^{\parallel}(\mathbf{p}, u))^*]} \tag{A21}$$

and for the perpendicular case as

$$D_S^{\perp}(\mathbf{p}, u) = \frac{-2\pi i m_D^2 T (1-v^2)^{3/2}(2 + v^2 - v^2 \sin^2(\theta) \cos^2(\phi - \beta))}{2p(1 - v^2 + v^2 \sin^2(\theta) \cos^2(\phi - \beta))^{5/2}[p^2 + \Pi_R^{\perp}(\mathbf{p}, u)][p^2 + (\Pi_R^{\perp}(\mathbf{p}, u))^*]}. \tag{A22}$$

The in-medium permittivity can now be calculated from

$$\varepsilon^{-1}(\mathbf{p}, u) = -\lim_{p_0 \to 0} p^2 D_{11}(p_0, \mathbf{p}, u), \tag{A23}$$

where $D_{11}$ is given in Eq. (A6). From Eq. (A17) it is seen that the real part is given by

$$\text{Re}\varepsilon^{-1}(\mathbf{p}, u) = -\text{Re}\left[\frac{p^2}{p^2 + \Pi_R(\mathbf{p}, u)}\right], \tag{A24}$$

where the retarded self-energies for the parallel and perpendicular cases are given in Eqs. (A14) and (A15), respectively. The imaginary part of the permittivity arises entirely from the symmetric propagator in Eqs. (A21) and (A22). One notices that other than the trigonometric angular factors, the momentum structure takes a very similar form to that given in Eq. (12), with the Debye mass being replaced by the retarded self-energies. Indeed it is easily checked that taking the ($v \to 0$) limit recovers the static expression.

With the entire framework now in place, we can apply our procedure of modeling the potential via the generalized Gauss law and linear response ansatz. This is again where our analysis takes a different path from the existing literature. As for the static case, our method requires solving two ordinary differential equations [Eqs. (10) and (11)] but now with a modified right-hand side that includes the in-medium finite-velocity complex permittivity. The first step is to ascertain the real-space expression $\varepsilon^{-1}(\mathbf{r}, u)$. This proves to be somewhat trickier than in the static case, due to the inherited angular dependence of the self-energies. For the parallel case, the integrand for the real part can be manipulated to

$$\text{Re}\varepsilon^{-1}(\mathbf{r}, u) \sim \int_0^{\pi/2} d\theta \int_{-\infty}^{\infty} dp \, p^2 \sin(\theta) h(\theta, u)$$

$$\times \text{Re}\left[\frac{p^2}{p^2 + \Pi_R(\mathbf{p}, u)}\right] e^{ipr\cos(\theta)}, \tag{A25}$$

where $h(\theta, u)$ contain the appropriate angular factors. A similar expression exists for the perpendicular case, with an extra integral over the azimuthal angle. Our attempts to compute this integral via the usual contour techniques led to an ill-defined limit. Thus we instead propose to follow the steps in [43]; that is, we solve for the Coulombic part of the potential in momentum space before Fourier transforming back to attain the real space expression. This is formally correct and does not conflict with any other part of our procedure. We then assume that the deduced $\text{Re}V_C$ satisfies our defining in-medium equation,

$$-\nabla^2 V_C(\mathbf{r}, u) = 4\pi\tilde{\alpha}_s \varepsilon^{-1}(\mathbf{r}, u), \tag{A26}$$

and apply the derivatives in order to acquire the in-medium permittivity to be used for the string part. This procedure is further justified since the resulting real part of the real-space permittivity reduces to the correct expression in the $v \to 0$ limit. The final expression for the parallel case is given in the main text [Eq. (39)], which leads to

$$\text{Re}\varepsilon^{-1}(\mathbf{r} \parallel \mathbf{v}) = \frac{1}{4\pi r} \int_0^{\pi/2} d\theta \sin(\theta) \cos(\theta)$$

$$\times \text{Re}\left[\sqrt{\Pi_R^{\parallel}(\theta, v)}\right]^2 e^{-\text{Re}\left[\sqrt{\Pi_R^{\parallel}(\theta, v)}\right] r\cos(\theta)}$$

$$\times \left(\text{Re}\left[\sqrt{\Pi_R^{\parallel}(\theta, v)}\right] r\cos(\theta) - 2\right). \tag{A27}$$

For the perpendicular case the results are identical after the replacements in Eqs. (41) and (42).

These expressions merit some discussion. First, we highlight that the potentials here differ slightly from those in [43]. In the computation of Eq. (39), one is faced with a momentum integral such as

$$\text{Re}V_C \sim \int d^3\mathbf{p} \, h(\theta, u) \text{Re}\left[\frac{1}{p^2 + \Pi_R(\theta)}\right], \tag{A28}$$

where the self-energy contains an angular dependence. The strategy is first to exploit a symmetry present in $\Pi_R$, namely that it is symmetric around $\theta = \pi/2$. After some manipulations this allows the $d_p$ integral to be extended over the entire real domain, such that a contour integration can be performed by continuing $p$ into the complex plane. One can see in the denominator in Eq. (A28) that the poles exist at $\pm i\sqrt{\Pi_R}$. Thus taking the real part—in accordance with the retarded propagator definition—must be done after the identification of the pole location. In [43] the taking of the real part was performed on the entire result of the contour integral, which can be seen in the discrepancy between Eqs. (34) and (35) in that text. We have also checked this numerically and confirmed that our expressions are correct. The second rather technical detail that arises from performing the contour integral is that closing in the upper-half plane gives a nonvanishing contribution. This leads to the term $\sim 1/r$ in Eq. (39) that is necessary to ensure the correct expression is recovered in the limit $v \to 0$.

The computation of the imaginary part of the real-space permittivity is somewhat simpler. It amounts to inverse Fourier transforming Eqs. (A21) and (A22) with an extra factor of $p^2$. The momentum integrals can be performed analytically, and the resulting expressions are angular integrals as we have just seen. In practice we have found that carrying out the same procedure as for the real part, i.e., computing $\mathrm{Im}\varepsilon^{-1}(\mathbf{r}, u)$ by taking appropriate derivatives of the existing $\mathrm{Im}V_C$ in [43], leads to an expression that gives the same result as the direct transform; however, it is numerically more stable and faster to evaluate. The final result for the parallel case is

$$\mathrm{Im}V_C(\mathbf{r} \parallel \mathbf{v}) = 2\tilde{\alpha}_s T \int_0^{\pi/2} d\theta \sin(\theta) \frac{(1-v^2)^{3/2}(2+v^2\sin^2(\theta))}{2(1-v^2\sin^2(\theta))^{5/2}} \frac{m_D^2}{2i\mathrm{Im}\Pi_R^{\parallel}(\theta, v)}$$
$$\times \left\{ \left[ \sinh\left(r\cos(\theta)\sqrt{\Pi_R^{\parallel}(\theta,v)}\right) \mathrm{Shi}\left(r\cos(\theta)\sqrt{\Pi_R^{\parallel}(\theta,v)}\right) \right. \right.$$
$$\left. - \cosh\left(r\cos(\theta)\sqrt{\Pi_R^{\parallel}(\theta,v)}\right) \mathrm{Chi}\left(r\cos(\theta)\sqrt{\Pi_R^{\parallel}(\theta,v)}\right) \right]$$
$$- \left[ \sinh\left(r\cos(\theta)\sqrt{\Pi_R^{\parallel}(\theta,v)^*}\right) \mathrm{Shi}\left(r\cos(\theta)\sqrt{\Pi_R^{\parallel}(\theta,v)^*}\right) \right.$$
$$\left. \left. + \cosh\left(r\cos(\theta)\sqrt{\Pi_R^{\parallel}(\theta,v)^*}\right) \mathrm{Chi}\left(r\cos(\theta)\sqrt{\Pi_R^{\parallel}(\theta,v)^*}\right) \right] \right\}, \quad (A29)$$

where $\mathrm{Shi}(x)$ and $\mathrm{Chi}(x)$ are defined, respectively, by

$$\mathrm{Chi}(x) = \gamma_E + \log(x) + \int_0^x dt \frac{\cosh(t) - 1}{t}, \quad (A30)$$

$$\mathrm{Shi}(x) = \int_0^x dt \frac{\sinh(t)}{t}. \quad (A31)$$

The perpendicular alignment is again obtained by the replacements in Eqs. (41) and (42). We do not include the corresponding expression for $\mathrm{Im}\varepsilon^{-1}(\mathbf{r}, u)$ since it is rather long and does not provide any intuition. With a computer algebra program, it can easily be obtained by acting the Laplacian on Eq. (A29) or the perpendicular case equivalent.

Finally, our expressions for the string part in-medium finite-velocity potential are then achieved in the same manner as for the static case, that is, via

$$V_S(\mathbf{r}, u) = c_0 + c_1 r - 4\pi\sigma \int_0^r dr' \int_0^{r'} dr'' r''^2 \varepsilon^{-1}(\mathbf{r}'', u). \quad (A32)$$